\documentclass[usenatbib]{mn2e}

\input psfig.tex

\usepackage{graphicx}
\usepackage{natbib}
\usepackage{array}
\bibpunct{(}{)}{;}{a}{}{,}

\usepackage{rotating}    % for sideways tables/figures

\input psfig.tex
\usepackage{latexsym}
\usepackage{natbib}
\usepackage{amssymb}
\usepackage{amsmath}

\usepackage{graphicx}
\usepackage{graphics}
\usepackage{fancyhdr}
\usepackage{morefloats}

\begin{document}

\title{The Origin of the Mass-Radius Relation of Early-Type Galaxies}

\author[C. Chiosi, E. Merlin and L. Piovan]{C. Chiosi$^{1}$\thanks{E-mail: cesare.chiosi@unipd.it},
E. Merlin$^{1}$ and L. Piovan$^{1}$ \\
$^{1}$Department of Astronomy, University of Padova, Vicolo dell'Osservatorio 3, 35122 Padova, Italy }

\date{Received: August 2012 ; Accepted:  }

\pagerange{\pageref{firstpage}--\pageref{lastpage}} \pubyear{2012}

\maketitle

\label{firstpage}

%=================== BEGIN ABSTRACT ===================

\begin{abstract} 

{Early-type galaxies obey a narrow relation traced by their stellar content between the mass and size (Mass-Radius relation). The wealth of recently acquired observational data essentially confirms the classical relations found by Burstein, Bender, Faber, and Nolthenius, i.e.  $\log R_{1/2} \propto \log M_{s}^{\simeq 0.54}$ for high mass galaxies and $\log R_{1/2} \propto \log M_{s}^{\simeq 0.3}$ for dwarf systems (shallower slope), where $R_{1/2}$ and $M_s$ are the half-light radius and total mass in stars, respectively. Why do galaxies follow these characteristic trends? What can they tell us about the process of galaxy formation?} {We investigate the mechanisms which concur to shape the Mass-Radius relation, in order to cast light on the physical origin of its slope, its tightness, and its zero point.} {We perform a theoretical analysis, and couple it with the results of numerical hydrodynamical (NB-TSPH) simulations of galaxy formation, and with a simulation of the Mass-Radius plane itself.} {We propose a novel interpretation of the Mass-Radius relation, which we claim to be the result of two complementary mechanisms: on one hand, the result of local physical processes, which fixes the ratio between masses and radii of individual objects; on the other hand, the action of cosmological global, statistical principles, which shape the distribution of objects in the plane. We reproduce the Mass-Radius relation with a simple numerical technique based on this view.}{} %{If our interpretation is correct, early-type galaxies formed at high redshifts via primordial mergers of small subunits, and fixed their dimensions \textit{ab initio} with little modifications in later times. Furthermore, most of them were formed \textit{before} $z \simeq 2 - 1$, thus ruling out the necessity for late mergers.}

\end{abstract}

\begin{keywords}
Galaxies, cosmology
\end{keywords}

%=================== BEGIN DOCUMENT ===================%

\section{Introduction} \label{intro}

Galaxies are known to exist in a variety of morphological types going from dwarf objects (in turn grouped in irregulars, ellipticals and spheroidals), to disk galaxies with the central bulges of different size (the origin of which is not yet clearly established), and finally to the elliptical ones of large mass and dimensions. The mass and size among the different types may vary by orders of magnitudes. This taxonomy implies a coherent picture of galaxy formation and evolution in cosmological context. Indeed, the situation observed in the local Universe is further complicated going back into the past at higher and higher redshifts, where a large variety of objects seems to exist.

\textbf{Galaxy formation in cosmological context}.
In the presently accepted model of the Universe, dominated by cold Dark Matter (DM) and by a mysterious form of energy expressed by the cosmological constant, and containing Baryonic Matter (BM) and photons (thereinafter $\Lambda$-CDM Universe), the cosmic structures are originated from the gravitational collapse of DM halos that gives rise to complexes of larger and larger scale within which baryons (gas) infall, forming stars and galaxies in a complicate game of star formation (SF), chemical enrichment, gas heating and cooling, and galactic winds. Many years of observations and theoretical speculation have clarified that building up galaxies of different morphological type (schematically irregulars, spirals and ellipticals) or even different parts of the same galaxy, e.g. the bulge, halo and disc in spirals, cannot be reduced to a unique physical mechanism. While there is sort general consensus on the formation of spiral and irregular galaxies, the question is unsettled in the case of early-type galaxies (ETG), the massive ones in particular. It is unclear how the $\Lambda$-CDM cosmological model should be reconciled with the observational claims for large, and red, galaxies already in place at very high redshifts \citep[e.g.][]{Harrison2011}. The problem is twofold: on one hand, it must be explained how and when star formation is quenched in both massive and small haloes, which is necessary to reconcile the theoretical prediction with the observed galaxy mass function \citep[see e.g.][]{Bundy2006}; on the other hand, one should also clarify how such massive systems as those recently observed can form at very high redshifts. Historically, two opposite scenarios exist:

(1) \textsf{the monolithic view}, which predicts that most of the mass in massive spheroids is already assembled in a strong and rapid burst of SF at high redshift. Galaxies evolve almost passively thereafter, without being strongly affected by subsequent mergers \citep[see][and references]{Chiosi2002,Peebles2002}. This picture does not conform to the present theoretical, concordance scenario of cosmology;

(2) \textsf{the hierarchical view}, according to which DM haloes (and BM inside) grow by mergers of smaller units in the dry (gas-poor, no SF) or wet (gas-rich and SF) mode \citep[see][and references therein]{DeLucia2006}. However the role played by mergers (major, minor, wet and dry) is still uncertain \citep{Hopkins_etal_2010} and several different versions of the same tell are around \citep[e.g.][]{Gonzalez_etal_2011}. Detailed numerical simulations have shed doubts on the coherence with which mergers can bring galaxies along the tight structural relations observed in the local Universe \citep[e.g.][]{Nipoti_etal_2009}.

\textsf{The early hierarchical or quasi monolithic view}. The modern picture, lately emerging from observational data and theoretical investigations, is as follows: at a certain (high) redshift, perturbations made of DM and BM detach themselves from the Hubble flow, and collapse on their own forming a proto-galaxy rich of sub-structures inside the common gravitational potential well. They merge, form stars and eventually give origin to a single galaxy. Depending on the initial density and/or angular momentum, the end product is different. Systems with low (or nearly zero) angular momentum will end up as spherical or elliptical objects. The massive ones and/or those originated by strong density perturbations had an early dominant episode of SF ever since followed by quiescence, whereas the low mass and/or lower density ones undergo a series of SF episodes along the Hubble time (\textit{galactic breathing}). This view is supported by the NB-TSPH simulations of \citet{Kawata1999,Kawata2001a,Kawata2001b}, \citet{KawataGibson2003a,KawataGibson2003b},
\citet{Kobayashi2005},  \citet{Merlin2006,Merlin2007}, and \citet{Merlin2012}. Likely objects with high angular momentum  end up as disc galaxies, in which the local low density environment favours SFs prolonged all over the galaxy life. The bulge (if present) likely follows the scheme envisaged for ETGs. Finally, irregular galaxies should originate from very low density perturbations within a shallow potential well undergoing intermittent and perhaps delayed SF. The advantage with this scenario is that ETGs in place at redshift greater than 2 are possible  \citep{Kobayashi2005,Merlin2006,Merlin2012}, and many structural and chemical properties are
explained \citep[see][]{Matteucci2007,Chiosi2007ASPC}.

\textbf{Masses and Radii of galaxies}.
In recent years, much attention has been paid to the Mass-Radius Relationship (MRR) of galaxies, in particular the ETGs and the compact and passive ones at high $z$. The MRR is indeed basic to any theory of galaxy formation and evolution.
The subject of the MRR of galaxies from ETGs to dwarf ellipticals and dwarf spheroidals, including also bulges and Globular Clusters has been recently reviewed by \citet{Graham2011} to whom we refer for many details.  The current MRR for ETGs will be presented below in great detail.

In addition to this, convincing evidence has been gathered that at relatively high redshifts, objects of mass comparable to that of nearby massive galaxies but smaller dimensions exist. These ``compact galaxies'' are found up to $z \geq 3$ with stellar masses from $10^{10}$ to $10^{12} M_\odot$ and half-light radii from 0.4 to 5 kpc (i.e. 3 to 4 times than more compact than the local counterparts of the same mass), and in nearly similar proportions there are galaxies with the same mass but a variety of dimensions \citep[e.g.,][]{Mancini_etal_2009,Valentinuzzi_etal_2010a}, and bulge to disk ratios \citep[e.g.,][]{Vanderwel2011}.
However, we will consider here only the case of standard ETGs, leaving the  compact galaxies aside but for a qualitative suggestion about their interpretation.

In the present study we ask ourselves the question: Why do ETGs obey a rather narrow mass-size relation instead of scattering around showing a broader combination of these two parameters? Spurred by this, we look for general physical principles governing this important scale relation. To clarify the aims and the methods of this study, we anticipate here the essence of our analysis. We speculate that the observed MRR for ETGs is the result of two complementary mechanisms. On one hand, the mass function of DM haloes hosting the visible galaxies gives (i) the typical cut-off mass at which, at any redshift, haloes become ``common'' on a chosen spatial scale, and (ii) the typical epoch at which low mass haloes begin to vanish at a rate higher than that at which they are born, because of merger events. On the other hand, these constraints define two loci (curves) on the MR-plane, because to each mass and formation redshift a typical dimension (i.e., radius) can be associated (using a basic relation between mass and radius of a collapsing object). If the typical dimension of a galaxy is somehow related to that of the hosting DM halo (as our NB-TSPH models seem to suggest), then the region of the MR-plane between the two limits fixed by the halo mass function is populated by galaxies whose dimensions are fixed at the epoch of formation, and only those objects that are ``possible'' at any given epoch may exist, populating a narrow region of the MR-plane.

The paper is subdivided as follows. In Section \ref{MRR} we present and discuss the observational MRR for a sample of ETGs galaxies taken from the SDSS catalogue. In Section \ref{models} we shortly describe the NB-TSPH models of ETGs that we used in the present analysis. In Section \ref{interp_MR} we present our interpretation of the MRR, based on elementary theories of cosmology and galaxy formation. In Section \ref{simul_MR} we simulate the MRR highlighting the deep causes governing the MRR. In Section \ref{compacts} we advance a possible explanation for the anomalous position of compact galaxies at high redshift. Finally, in Section \ref{conclusions} we draw some remarks and conclusions.

%%%%%%%%%%%%%%%%%%%%%%%%Fig 1
\begin{figure}
\centering{
\includegraphics[width=9.0cm,height=8.0cm]{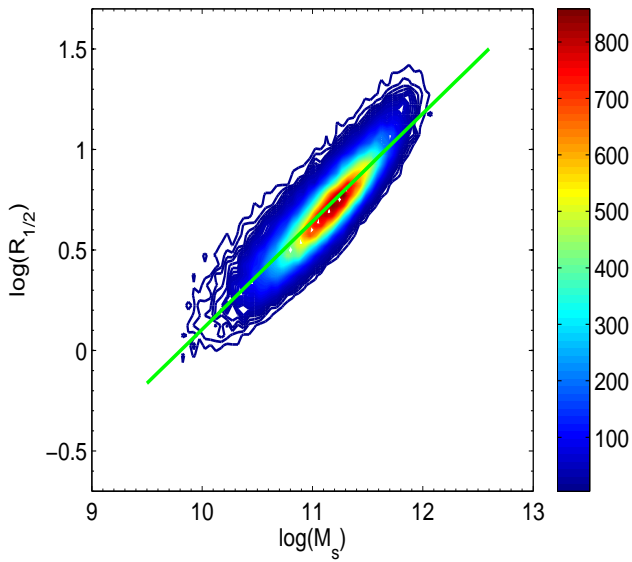}}
\caption{The $\log[R_{1/2}]$ versus $\log[M_{s}]$ relation for galaxies, where $R_{1/2}$ is the half-mass radius, and
$\log M_{s}$ is the total stellar mass. Although $R_{1/2}$ is not strictly identical to the effective radius $R_e$, they are very close to each other. Throughout this paper we will always use $R_{1/2}$, which is easier to calculate for dynamical NB-TSPH models of galaxies, and assume $R_{1/2}\approxeq R_e$. The contour lines display the regions of the MR-plane populated by the same number of objects (the bar yields the correspondence between color-code and number of galaxies).  The data are from the HB sample of \citet{Bernardi2010}. The green solid line is the linear best fit of eqn.(\ref{mr1}). }
\label{reffmass}
\end{figure}

\section{The observational Mass-Radius relation} \label{MRR}

An impressive body of data have been acquired concerning the masses and the dimensions of galaxies. %not only in the local old Universe but also in the distant and young one, thus making it feasible to address the question whether these two important parameters changed with time as predicted by any hierarchical mode of galaxy formation. 
The situation has been recently discussed by \citet[][]{Shankar_etal_2011} and \citet{Bernardi_etal_2011} and references therein.

\textsf{The normal early-type galaxies}. The observational data we are considering is the HB sample selected by \citet{Bernardi2010} from the SDSS catalogue, containing $\simeq 60,000$ galaxies\footnote{The selection conditions are $fracde[v_r]$=1 and $b/a> 0.6$, therefore the sample is dominated by elliptical galaxies \citep[see][for all details]{Bernardi2010}.}. The observational MRR is displayed in Fig. \ref{reffmass}. The linear best fit of the  SDSS data is

\begin{equation}
\log R_{1/2}=0.54\,\,\log M_{s} - 5.25 \label{mr1}
\end{equation}

\noindent where  $M_{s}$ in $M_\odot$ is the estimated stellar mass (in $M_{\odot}$) and $R_{1/2}$ (in kpc) is the radius containing half of it (nearly identical to the classical effective radius $R_e$). The slope (and zero point) of the above MRR is quite robust as it coincides with previous determinations: using the \citet{Burstein1997} data, the same slope has been found by \citet{Chiosi2002}, and using the SDSS data by \citet{Shenetal2003}.

The distribution of the bulk of galaxies is also confirmed by the smaller sample of \citet[][]{Shankar_etal_2011} always extracted from the SDSS survey but using slightly different selection criteria. The area covered by the observational data is slightly larger than that the \citet{Bernardi2010} data. In any case no significant differences can be noted.
Our analysis of the MRR will stand only on SDSS sample, thus securing homogeneity of mass and radius estimates. However to complete the scene it is worth looking at the position on the MR-plane of dwarf  galaxies. Since
no attempt is made to homogenize the data for these latter objects with those of the SDSS sample, dwarf  galaxies will not be included in the analysis of the MRR and will be considered separately.

\textsf{The dwarf galaxies}. The dwarf galaxies are taken from \citet{Burstein1997}. The sample is made of dEs and dSphs. It is worth recalling that the masses used by \citet{Burstein1997} are the dynamical masses and not the stellar masses, so this group is not strictly homogeneous with the sample for ETGs.  In addition to the few dwarfs of the  \citet{Burstein1997} sample, to complete the picture, we also consider the mean relationship for the dwarf galaxies of the Local Group according to the measurements made
by \citet{Woo2008} who yield the relationship

\begin{equation}
\log R_{1/2}=0.3\,\log M_{s} -2.7. \label{mrdwarf}
\end{equation}
to which also the \citet{Burstein1997} dwarfs seem to obey.

\textsf{Changing slope of the MRR}. It is soon evident that there is no unique slope for the MRR of the different groups of objects. The slope is 0.54 for ETGs and 0.3 and lower for  dwarf galaxies. Looking at the data in detail,
the slope is even steeper than 0.54 in the region of the largest and most massive ETGs going up to 1 and even more, see the top part of the MRR by \citet[][]{Bernardi2010}, \citet{Guo2009},  \citet{vanDokkum2010}, and Fig. 1 of \citet{Graham2011}. This is a point to keep in mind when interpreting the observational data.

\textsf{General Remarks}.
Information and details on how the stellar masses $M_s$, and half-mass radii, $R_{1/2}$, have been derived can be found in the original sources to which the reader should refer. Of course some possible systematic biases among the different sets of data are to be expected, whose entity, however, ought to be small. This is somewhat sustained by the overall agreement among different sources as far as some general  relationships are concerned, e.g.  the agreement in the slope of the MRR for ETGs between \citet{Bernardi2010} and \citet{Burstein1997}. The same for the dwarf galaxies. Perhaps somewhat  larger uncertainties are present in the case of   compact galaxies the data of which will be presented in Section \ref{compacts}. However, since the three groups of objects will be treated separately and only from a general qualitative point of view, no homogenization of the data is needed. Our discussion is limited to the SDSS sample, the data of which is not only internally consistent but also homogeneous to those of the theoretical simulations. As far as compact galaxies is concerned, what will matter here is the general consensus  that galaxies of very different dimensions and comparable stellar masses are found  \citep[e.g. see][for general discussion of the subject]{Shankar_etal_2011}.

\begin{figure}
\centering{
\includegraphics[width=.4\textwidth]{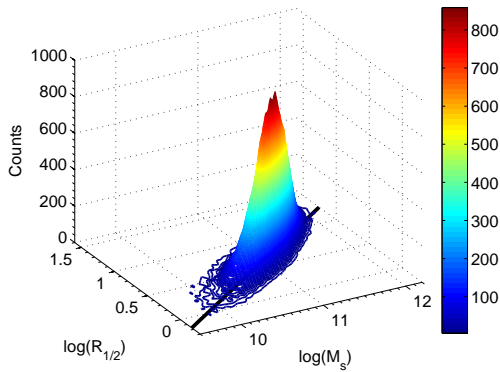}}
\caption{3D view of the MRR. The number frequency distribution of galaxies on the MR-plane: $M_{s}$ is the total stellar mass in solar units and  $R_{1/2}$ is the half-mass radius in kpc, and  ``Counts'' is number of galaxies falling within each cell of the plane with dimensions $\Delta \log M_{12}$=0.05 and $\Delta \log R_{1/2}$=0.05. The data are from the HB sample of \citet{Bernardi2010}. The thick line on the MR-plane is the linear fit of the data: $\log R_{1/2}=0.54\,\, \log M_{s} - 5.25$. }
\label{mass_rad_number}
\end{figure}

\textsf{Galaxy counts in the MR-plane}. Before proceeding further it is worth looking at the number frequency distribution of galaxies with given mass and radius. We limit ourselves to the SDSS sample. To this aim we divide the MR-plane in a grid of square cells with dimensions of 0.05 in units of $\Delta \log M_{s}/(10^{12}\,M_\odot)$ and $\Delta \log R_{1/2}$, and count the galaxies falling in each cell.  In Fig. \ref{mass_rad_number} we display the 3D space of these parameters to simultaneously highlight the distributions of the MRR along the direction parallel to the  best-fit line on the MR-plane and the direction perpendicular to it. The view angle is chosen in such a way that the projections on the plane parallel and perpendicular to the MRR line can be easily figured out. The projection on the  MR-plane has already been displayed in Fig. \ref{reffmass}.  While the projection on the plane parallel to direction of the best-fit line  can be easily understood in terms of selection effects (galaxy mass fall-off at the high mass end and lack of data at the low mass end), the distribution perpendicular to this   is more difficult to explain. Indeed \textit{galaxies tend to fall in a rather narrow strip of the MR-plane, tightly gathering around the line with slope 0.54}.

%%%%%%%%%%%%%%%%%%%%%%%%%%%Fig 2

\section{NB-TSPH models of galaxies on the MR-plane}\label{models}

Although a great deal of our analysis will be made using analytical relationships, some NB-TSPH models of ETGs by the authors are also considered. The models are presented in two groups: the so-called reference models by \citet{Merlin2012} which are calculated either to the present or down to redshift lower than z=1, and the ancillary models that are explicitly calculated for this paper and limited to initial stage. In Appendix A we briefly report on a few salient properties characterizing all  the models in use, whereas the complete description of the reference models and underlying evolutionary code can be found in \citet{Merlin2010,Merlin2012}.

For the sake of completeness, in this section we limit ourselves:

(i) to remind the cosmological scenario adopted to calculate the  galaxy models, i.e. the standard $\Lambda$-CDM model with $H_0$=70.1 km/s/Mpc, flat geometry, $\Omega_{\Lambda}$=0.721, $\sigma_8$=0.817 and the baryonic fraction $\simeq 0.1656$. The initial conditions (positions and velocities of the DM and BM particles) are taken from large scale cosmological simulations taking into account the expansion of Universe;

\begin{figure}%%%[htbp]
\centering{
\includegraphics[width=9.0 cm,height=8.0 cm]{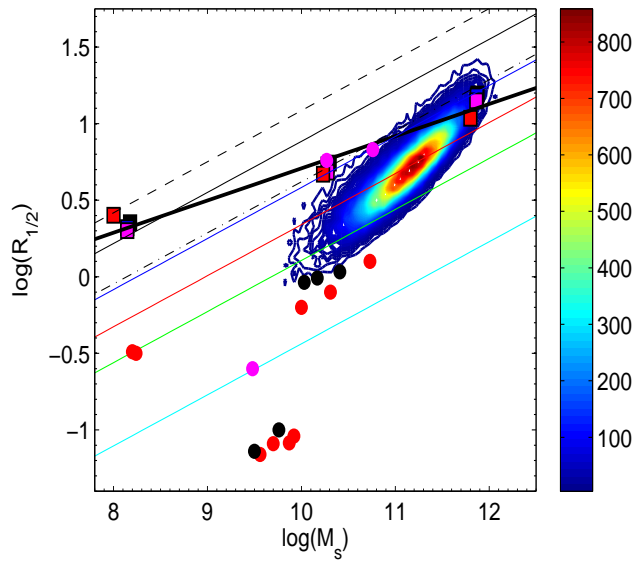}
\caption{Theoretical MRR, i.e. $\log R_{1/2}$ versus $\log M_s$, where $R_{1/2}$ is the half-mass radius and $\log M_s$ the total stellar mass. The filled squares are the twelve reference models of Table \ref{tab1}
and the circles are the ancillary models of Table \ref{tab2} of Appendix A.  The color code adopted for the various  models is in principle related to the initial over-densities. Since the entries of Tables \ref{tab1} and \ref{tab2} are fully sufficient to identify the models, the correspondence color-code-initial density is superfluous here. The thick black line is the fit of the twelve reference models. The thin straight lines are the theoretical MRR on which proto-galaxies of given DM over-density settle down at the collapse stage as a function of the redshift. The lines are given by eqn. (\ref{mr3}) taken from \citet{Fan2010}: solid lines are for $m=10$ and $f_\sigma=1$; the color indicates  different redshifts (black: $z=0$; blue: $z=1$; red: $z=2.5$; green: $z=5$;
cyan: $z=20$); the thin  black dashed line is for $m=10$, $f_\sigma=1.5$, and $z=0$;  Finally, the thin black dashed-dotted line is for $m=5$ and $f_\sigma=1$, and $z=0$.} }  \label{theo_re_mass}
\end{figure}

(ii) to display in Fig. \ref{theo_re_mass}  the position of the reference and ancillary models on the MR-plane;

(iii) to discuss two general aspects of the models that are relevant here, the \textit{filiation thread} and the \textit{paths of the models in the MR-plane}. \\

%%%%%%%%%%%%%%%%%%%%%%%Fig  3

\textsf{The filiation thread}. The models we have calculated at varying the initial cosmological density contrast $\delta\rho_i(z)$ and the star formation efficiency $\epsilon_{sf}$ help us to understand the different role played
by $\delta\rho_i(z)$ and the gas density $\rho_{g,sf}$ at the onset of star formation  in determining  the size of the resulting galaxy made of stars. Given the total mass oh the proto-halo $M_{DM}$, the cosmological density $\delta\rho_i(z)$ determines the initial radial dimension of the DM perturbation. This does not coincide with the initial radius at which star formation begins in the baryonic component of a galaxy. In other words, within the potential well of DM, the gas keeps cooling at increasing density and only when the threshold density for star formation is met (which in turn is related to   $\epsilon_{sf}$),  stars appear: the galaxy is  detectable on the MR-plane. In this latter step of the filiation thread, $\rho_{g,sf}$ plays the dominant role. The remaining gas continues to fall into the gravitational potential well until either it is exhausted by SF or it is expelled via shocks because of energy feed back. Thanks to this there will be a correlation between the DM and the final mass in stars, measured by $m= M_{DM}/M_s$ (with $m\sim 10$ on the average, see below). NB-TSPH models \citep{Merlin2012} indicate that the transformation of BM into stars occurs under the homology condition $G M_{DM}/R_{DM} \simeq G M_s/R_s$, i.e.  equal gravitational potential energy per unit mass of the two components.  In general, the model galaxies follow this rule and the two components of a galaxy lay on nearly parallel MRRs: i.e. $R_{DM}\propto M_{DM}^{1/3}$ and $R_s\propto M_s^{1/\beta}$ with $\beta\simeq 3$. At decreasing total mass, the exponent $1/\beta$ goes from 0.333 for galaxies with total mass $10^{13}\, M_\odot$ to about 0.2 for a mass of $10^9\, M_\odot$ or even lower. This deviation from the $R_s\propto M_s^{1/3}$ law can be  interpreted as due to an increasing departure from the condition of an ideal  collapse    because dissipative processes are now at work. The higher the initial mass, the closer the evolution of the proto-galaxy is to the simple collapse models. In other words, the straight collapse configuration corresponds to a minimum total energy of the system, whereas in all other cases the total energy system is far from the minimum level. Recasting the concept in a different way, the straight collapse is favored with respect to other energy costing configurations. Real galaxies tend to follow the rule by as much as they can compatibly with their  physical conditions (total mass, initial density, star formation history, ...). The result of it will be that the final model galaxies will be located on a new line (not necessarily a straight line but likely close to it) with a certain slope (about 0.2 in our case, see the linear fit of the final mean location of the  reference models). The slope is flatter than the 0.33 slope of the iso-density line of the initial DM proto-galaxies. There is not reason for the two slopes being the same.\footnote{In principle, the different redshift of the haloes at the collapse stage should be taken into account when considering the slope of their MRR. However, as our models  have quite similar collapse redshifts, $z\simeq 2.5 - 5$, the slopes of their MRRs do not change significantly.}

\textsf{Path of the models in the MR-plane}. Given these premises, a key question to clarify from the very beginning is the following: by how much our model galaxies, in which the so-called early hierarchical scheme is at work -- early aggregation of lumps of DM and BM, conversion of gas into stars according to some star formation history, expulsion of gas by galactic wind over long periods of time -- do change their star mass and dimensions during the whole evolutionary history? To this aim we show in Fig. \ref{reffmass1} the paths of some galaxy models per mass group
(all the others  have similar behavior). The circles of increasing size mark the stage at which the models have assembled $25\%$, $50\%$ and $75\%$ of the final stellar mass. Models move erratically in the MR-plane. The situation can be understood considering the definition of radius we have adopted: the galacto-centric distance inside which half of the stellar mass in enclosed.  The radius for the early evolutionary stages is rather uncertain. From an operational point of view, in principle at each evolutionary stage, once identified the barycenter of all the star particles existing at that time,  spheres of increasing radius centered at this point are drawn up to encompassing half of the stellar mass. In  the early evolutionary stages, characterized by captures of lumps of baryonic mass onto the already existing stellar body, desegregation of part of it by close encounters of other lumps of matter and similar events,  the barycenter and the size of the sphere containing half star mass are quickly changing with time. The situation stabilizes as more and more baryonic mass is turned into stars and the main body of the galaxy is shaped. All this reflects on the the quite erratic path of the models on the MR-plane. However, looking at the large-scale trend,  over the whole lifetime of the order of 13 Gyr, the star mass can increase by a factor of 10, whereas the radius remains nearly constant for the low and the intermediate mass galaxies. The situation is more confused for the high mass models.  In any case, each model galaxy wonders around in a rather small box of the MR-plane.

In this context we point out that, while $\epsilon_{sf}=1$ may concur to create diffuse stellar systems rather than compact objects because of the artificially high efficiency in peripheral regions, the physical argument behind the result is robust. The process of star formation most likely  depends on the local properties of the medium, primarily in terms of gas density, which (at least as a first approximation) depends on the depth of the potential well. Therefore, a ``threshold density'' (whether universal or not is still an open question) should be considered as an unavoidable ingredient in the scenario. Threshold density and efficiency parameter are unknown. However, they can vary without changing  the physical core of our conclusions: the position of a galaxy on the MR-plane is essentially decided in the early stages of its evolution.

%%%%%%%%%%%%%%%%%%%%%%%Fig 4
\begin{figure}
\centering{
\includegraphics[width=.45\textwidth]{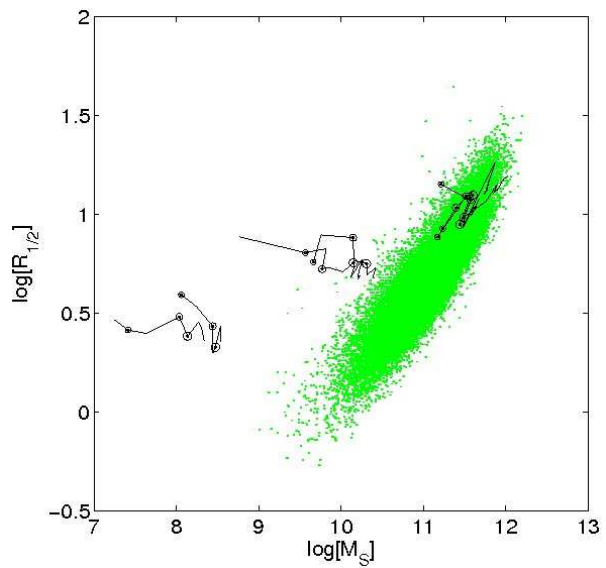}}
\caption{Path on the MR-plane of three reference models of different mass and same initail over-density. The open circles along the solid lines show the position  at the time at which $25\%$, $50\%$ and $75\%$ of their stellar content is assembled; the size of the circles is proportional to the assembled mass. As the stellar mass of the model galaxies grows monotonically with time, the path on the MR-plane is in general from  left to  right. However, some erratic displacements are possible owing to the definition itself of $R_{1/2}$, in particular during the earliest evolutionary stages.} \label{reffmass1}
\end{figure}

\section{Interpretation of the MR-plane}\label{interp_MR}

If we compare the present-day position of the reference NB-TSPH models on the MR-plane with the region populated by real galaxies (Fig. \ref{theo_re_mass}), at a first glance one would be tempted to conclude that only the high mass models fairly agree with observations, whereas the low mass ones (and to some extent also those of intermediate mass) apparently have too large radii with respect to their masses. However, before drawing the conclusion that essentially the models fail to reproduce the data, it is worth recalling that similar trend is first predicted by the filiation thread, second it has been found by other NB-TSPH  calculations,  \citet[see for instance][]{Chiosi2002}.

Is the observational sequence populated only by galaxies behaving as our massive ones? Or what else? Since for any value of the halo mass there is a certain redshift below which haloes of this mass start decreasing in number by mergers \citep{Lukic2007}, galaxies generated by those haloes become more and more unlikely as it should be the case for our low mass models. Indeed when at a given   redshift we have assumed the existence of haloes of any mass, we have neglected this important effect. Similar considerations would also apply to haloes of large mass. Therefore, the situation may occur that haloes/baryonic galaxies are calculated and plotted onto the MR-plane even though according to the above arguments their existence is very unlikely. On this ground, in the following we argue that the observational MRR of ETGs (galaxies in general) is the result of convolving two agents: the halo growth function providing the number density of haloes of different mass  as a function of the redshift (the concordance $\Lambda$-CDM Universe), and the fundamental MRR determining the size of a galaxy as a function of its mass and formation redshift. These two important ingredients are shortly presented below.

%%%%%%%%%%%%%%%%%%%%%Fig 5
\begin{figure}
\centering{
\includegraphics[width=9.0cm,height=8.0cm]{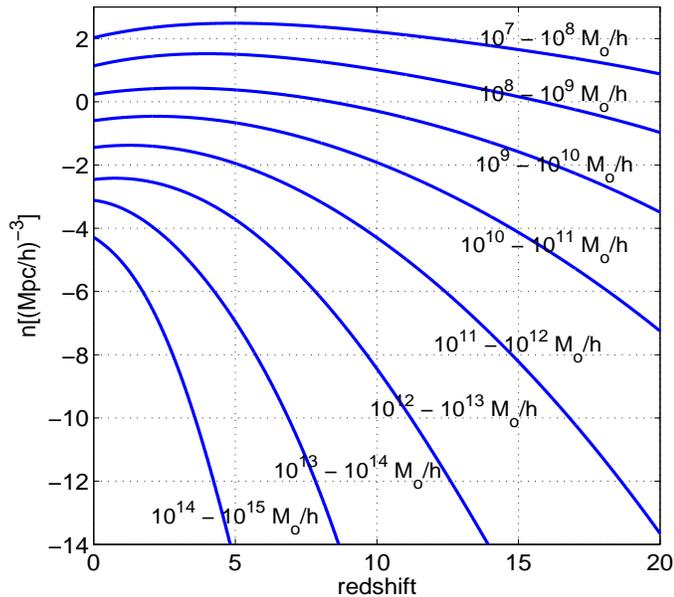}
\caption{The growth function of haloes $n(M_{DM}, z)$ reproduced from \citet{Lukic2007}. }  \label{lukic}}
\end{figure}

\subsection{ The halo growth function $n(M_{DM},z)$}

The distribution of the DM halo masses and their relative number density as a function of the redshift has been recently studied, among the others, by \citet{Lukic2007} who, using  the $\Lambda$-CDM cosmological scenario and the \citet[][]{Warren2006} mass function of haloes, derive the \textit{halo growth function} $n(M_{DM}, z)$. This gives the number density of haloes of different masses  per (Mpc/$h)^3$ resulting by all creation/destruction events. The growth function is expressed  in terms of the normalized Hubble constant $h=H_0/100$, where $H_0$ is assumed to be
$H_0$=70.1 Km/s/Mpc. The explored interval of redshift goes from 0 to 20. The  $n(M_{DM}, z)$ function of \citet{Lukic2007} is shown in Fig. \ref{lukic}\footnote{This has been derived from an analytical interpolation of the  data presented in Fig. 1 of \citet{Lukic2007}. More details are given below.}.

Although what we are going to say is well known, see the pioneer study of \citet{Press1974} and \citet[][ for ample referencing]{Lukic2007}, for the sake of clarity and as relevant to our discussion we note the following: (i) for each halo mass (or mass interval) the number density is small at high redshift, increases to high values toward the present, and depending on the halo mass either gets a maximum value at a certain redshift followed by a decrease (typical of low mass haloes) or it keeps increasing as in the case of high mass haloes; in other words, first creation of haloes of a given mass (by spontaneous growth of perturbation to the collapse regime or by mergers) overwhelms their destruction (by mergers), whereas the opposite occurs past a certain value of the redshift, for low mass halos; (ii) at any redshift high mass haloes are orders of magnitude less frequent than the low mass ones; (iii) at any redshift, the mass distribution of haloes has a typical interval of existence whose upper mass end (cut-off mass) increases at decreasing redshift.

\subsection{The fundamental Mass-Radius Relationship}

Spherical DM perturbations (with BM inside) that  undergo collapse when the density contrast with respect to the surrounding medium  reaches a suitable value obey the  MRR  given by

\begin{equation}
\left({4\pi \over 3} \right) R_{DM}^3 = {M_{DM} \over \lambda
\rho_u(z)} \label{mr2}
\end{equation}

\noindent where $\rho_u(z)$ % \propto (1+z)^3$
is the density of the Universe at the redshift $z$, and $\lambda$ is the factor for  the density contrast of the DM halo. This expression is of general validity whereas the  function $\lambda$  depends on the cosmological model of the Universe, including the $\Lambda$-CDM case. All details and demonstration of it can be found in \citep[][ their Eq. 6]{Bryan1998}.

In the context of the $\Lambda$-CDM cosmology, \citet{Fan2010} have adapted the general relation (\ref{mr2}) to provide an expression  correlating the halo mass $M_{DM}$ and the  star mass $M_s$ of the galaxy born inside it,  the half light (mass) radius $R_{1/2}$ of the stellar component, the redshift at which the collapse takes place $z_f$, the shape of the BM galaxy via a coefficient $S_S(n_S)$ related to the Sersic brightness profile from which the half-light radius is inferred and the Sersic index  $n_S$,  the velocity dispersion of the BM component with respect to the that of DM (expressed by the parameter $f_\sigma$), and finally the ratio  $m=M_{DM}/M_S$. The expression  is

\begin{equation}
R_{1/2}=0.9 \frac{S_S(n)}{0.34}\frac{25}{m} \left( \frac{1.5}{f_\sigma} \right)^2 \left( \frac{M_{DM}}{10^{12} M_\odot} \right)^{1/3} \frac{4}{(1+z_{f})}.  \label{mr3}
\end{equation}

\noindent Typical value for the coefficient $S_S(n_S)$ is 0.34. For the ratio $m= M_{DM}/M_s$ the empirical data confine it in the range 20 to 40, whereas our NB-TSP models yield  $m\simeq 15$ to 20 depending on the particular case we refer to.  For the purposes of the present study we round the ratio to 10 (see below). Finally $f_\sigma$ yields the three dimensional star velocity dispersion as a function of the DM velocity dispersion, $\sigma_s=f_\sigma \sigma_{DM}$. Here we adopt $f_{\sigma}=1$.  For more details see \citet{Fan2010} and references therein.

We point out that this relation is strictly valid only for monolithic infall of BM into collapsing DM potential wells. Therefore, its application is not justified when dealing with late mergers between already virialized objects, or with different mechanisms of mass assembly, such as, in particular, an early-hierarchical scenario. Nevertheless, this formula provides a general reference to obtain the typical dimension of a galactic system as a function of its mass and formation redshift. While adjustments are possible, the general trend is well defined. However, some deviations from this law are possible and expected, e.g. for low redshifts. See below for further discussion.

The slope of relation (\ref{mr3}) is nearly identical to (only slightly steeper than) the slope estimated from the NB-TSPH models; the difference can be fully ascribed to the complex baryon physics, which causes the stellar system to be slightly offset with respect to the locus analytically predicted from DM haloes. Therefore, a  model slope (close to 1/3)  different from that of the observational MRR is not the result of inaccurate description of the physical processes taking place in a galaxy; on the contrary, it mirrors the fundamental relationship between mass and radius in any system of given mean density. Indeed it is remarkable that quite complicated numerical calculations clearly display this fundamental feature. \textit{If this is the case, why do real galaxies gather along a line with a different slope?}

%%%%%%%%%%%%%%%%%%%%%Fig 6
\begin{figure}
\centering{
\includegraphics[width=9.0cm,height=8.0cm]{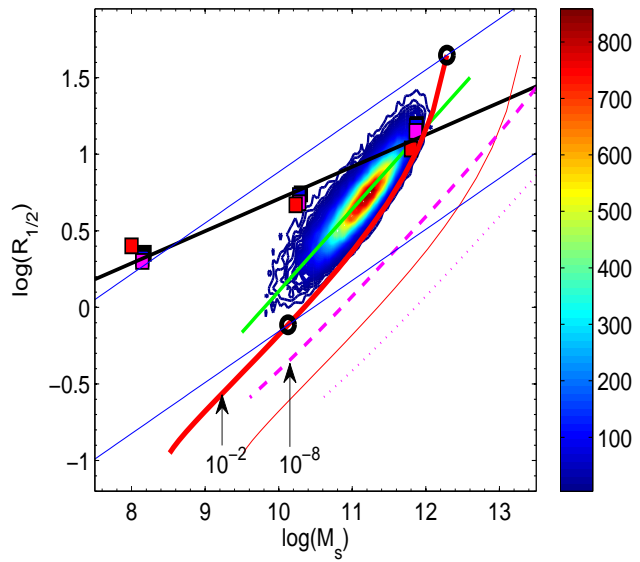}}
\caption{The \textit{cosmic galaxy shepherd} (CGS) and the corresponding locus of DM parent haloes (the red thick and thin solid lines, respectively). The ``statistics'' is $10^{-2}$ haloes per (Mpc/h)$^3$. In addition to this we show the case with ``statistics'' of  $10^{-8}$ haloes per (Mpc/h)$^3$ (the magenta thick and thin dashed lines). The various loci are plotted onto the observational MR-plane (see text for all details). We also draw the observational data for the  HB sample of \citet{Bernardi2010} with their linear fit (solid black line),  and two theoretical MRR from eqn. (\ref{mr3}) by \citet{Fan2010} (solid thin blue lines), relative to $z=0$ and $z=10$, with $m=10$ and $f_\sigma=1$. The present day position of the reference galaxy models and their linear fit are shown (filled squares and black solid line). Finally, the filled black circles show the intersection of the CGS with the theoretical MRR for the case with $10^{-2}$ haloes per (Mpc$/h)^3$.} \label{reffmass1}
\end{figure}

\subsection{The cosmic galaxy shepherd}

To answer the above question, we attack the problem from a different perspective, trying to investigate whether the observational MRR owes its origin to deeper reasons, likely related to the growth function of DM haloes  \citep{Lukic2007}.

Given a certain fixed number density of haloes $N_s$ (thereinafter referred to as a ``statistics''), on the $n(M_{DM}, z)-z$ plane of Fig. \ref{lukic} this would correspond to an horizontal line intersecting the curves for the various  masses at different redshifts, i.e. obeying the equation $n(M_{DM}, z)= N_s$. Each intersection provides a pair ($M_{DM}$, $z$) which gives the mass of the haloes fulfilling the statistics $N_s$ at the  corresponding redshift $z$ (or viceversa the redshift satisfying the statistics for each halo mass). For any value $N_s$ we get an array of pairs ($M_{DM}$, $z$) that can be extrapolated to a continuous function that,  with aid the \citet{Fan2010} relationship (in which the parameters $m$ and $f_\sigma$ are fixed), provides the  corresponding relationship between the mass in stars and the half-mass radius  of the baryonic galaxy associated to a generic host halo to be plotted on the MR-plane.

Repeating the procedure for different values of $N_s$, we get a manifold of curves on the MR-plane. It turns out that with the ``statistics'' corresponding to $10^{-2}$ haloes per (Mpc$/h)^3$, the curve is just at the edge of the observed distribution. Higher values of the statistics would shift it to larger haloes (baryonic galaxies), the opposite for lower values of the ``statistics".  Why is $N_s=10^{-2}$ haloes per (Mpc$/h)^3$ so special? Basing on crude,  simple-minded arguments we recall that the total number of galaxies observed by the SDSS amounts to about $\simeq 10^6$, whereas the volume of Universe covered by it is about $\simeq 1/4$ of the whole sky times a depth of $\simeq 1.5\times 10^9$ light years, i.e. $\simeq 10^8$ Mpc$^3$, to which the number density of about $10^{-2}$ haloes per (Mpc$/h)^3$ would correspond\footnote{We are well aware that this  is a very crude estimate not taking into account many selection effects both in the observations and the halo statistics based on NB simulations, such as  the \citet{Lukic2007} plane itself. However, just for the sake of argument, we can consider it  as a good estimate to start with.}.

The analytical fit of this curve is

\begin{flalign}
   \log R_{1/2} =&  0.048562 (\log M_s)^3 -1.4329 (\log M_s)^2 \nonumber \\
                 &+ 14.544 (\log M_s) -50.898 \label{shepherd_rel}
 \end{flalign}

\noindent and we note that the slope gradually changes from 0.5 to 1 and above as we move from the low mass to the high mass range. It is worth recalling here that a similar trend for the slope is also indicated by the observational data \citep[see][and references therein]{vanDokkum2010}. Owing to the many uncertainties we do not try to formally fit the median of the empirical MRR, but we limit ourselves to show that the locus predicted by the  $10^{-2}$ haloes per (Mpc$/h)^3$ ``statistics'' falls on the MR-plane close to the observational MRR. Lower or higher values of the halo number density would predict loci in the MR-plane too far from the observational MRR unless  other parameters of eqn. (\ref{mr3}), e.g.  $m$, are drastically changed  assuming values that are difficult to justify.

Finally, we call attention on the fact the locus on the MR-plane defined by the relation (\ref{shepherd_rel}) is ultimately related to the top end of the mass scale of haloes (and their filiated baryonic objects) that can exist at each redshift. In other words, recalling that the mass of any intersection pair for $N_s=10^{-2}$ corresponds to haloes becoming statistically significant in number on the observed spatial scale at the associated redshift, this can be interpreted as the so-called cut-off mass in the \citet{Press1974} or equivalent formalisms \citep[see][ for details and references]{Lukic2007}.
Therefore, this provides also an upper boundary to the mass of galaxies that are allowed to be in place (to collapse) at each redshift. We name these locus the {\it Cosmic Galaxy Shepherd} (hereafter CGS). All this is shown in Fig. \ref{reffmass1}, where we also plot the curves relative to another possible ``statistics'' -- that is, $10^{-8}$ haloes per (Mpc$/h)^3$, corresponding to 1 halo per $10^8$ (Mpc$/h)^3$, for the sake of comparison.

There are two points to be clarified. First, this way of proceeding implies that \textit{each halo hosts one and only one galaxy and that this galaxy is an early type object matching the selection criteria of the \citet{Bernardi2010} sample}. In reality ETGs are often seen in clusters and/or groups of galaxies and many large spirals are present. Only a fraction of the total population are  ETGs. One could try to correct for this issue by introducing some empirical statistics about the percentage of ETGs among all types of galaxy. Despite these considerations, to keep the problem simple we ignore all this and stand on the minimal assumption that each DM halo host at least one baryonic component made of stars. This is a strong assumption, on which we will come back again later on. Second, we have assumed $m=10$ and $f_{\sigma}=1$. According to \citet{Fan2010}, the empirical estimate of $M_{DM}/M_s$ ration is about 20-40, i.e. a factor of two to four less efficient star formation than we have assumed basing on our NB-TSPH models. However, a smaller value for $m$ does not invalidate our analysis, because it would simply shift the location of the baryonic component on the MR-plane corresponding to a given halo ``statistics''. Finally, $f_\sigma=1$ is a conservative choice. The same considerations made for $m$ apply also to this parameter. At present, there is no need for other values.

Along the CGS, redshift and cut-off mass go in inverse order, i.e. low masses (and hence small radii) at high redshift and viceversa. This means that a manifold of MRRs defined by eqn. (\ref{mr3}), each of which referring to a different collapse redshift, can be  selected, and along each MRR only masses (both parent $M_{DM}$ and daughter $M_s$) smaller than the top end are permitted, however each of which with a different occurrence probability: low mass haloes are always more common than the high mass ones. In the observational data,  it looks as if ETGs should occur only towards the high mass end of each MRR, i.e. along the locus on the MR-plane whose right hand side is limited by the CGS. This could be the result of  selection effects, i.e. (i) galaxies appear as ETGs only in a certain interval of mass and dimension and outside this interval they appear as objects of different type (spirals, irregulars, dwarfs etc..), or (ii) they cannot even form or be detected (e.g. very extended objects of moderate/low mass). Finally, in addition to this, we  argue that  another physical reason limits the domain of galaxy occurrence also on the side of the low mass, small dimension objects. We will come to this later on.\\

\textsf{Dissipation-less Collapse}.
It is an easy matter to understand that the CGS is another way of rephrasing the top-hat spherical dissipation-less collapse for primordial fluctuations by \citet{Gott_Rees1975}. For the sake of easy understanding we summarise the key steps of the dissipation-less collapse using the version given by \citet{Faber1984} and \citet{Burstein1997}. Let $\delta$ be the rms amplitude of primordial density perturbations

\begin{equation}
      \delta \propto   M^{-{1 \over 2} - {n \over 6} }
\label{delta_M}
\end{equation}
\noindent where $M$ is the mass at the initial red-shift, and $n$ is the slope of the density fluctuation $\delta$. After collapse, the equilibrium structure of a DM halo originated from given $\delta$ and M  follows the relations \citep{Gott_Rees1975}

\begin{equation}
      R   \propto   \delta^{-1} M^{1 \over 3}
\label{delta_MR1}
\end{equation}
\noindent from which we immediately get

\begin{equation}
      R   \propto M^{ {5+n \over 6}}
\label{delta_MR2}
\end{equation}
\noindent As already pointed out by \citet{Burstein1997}, inserting $n=-1.8$, the power spectrum of CDM \citep{Blumenthal1984}, we get the  relation

\begin{equation}
      R_{DM}   \propto M_{DM}^{0.53} \
\label{delta_MR3}
\end{equation}
\noindent The slope of the MRR derived from the dissipation-less collapse is the same of eqn. (\ref{mr1}) all over the mass range from normal/giant galaxies, M32 and $\omega$Cen like objects, to classical Globular Clusters. This was already pointed out long ago by \citet{Chiosi2002} and discussed by \citet{Graham2011}.  It can be easily checked by   plotting the data for Globular Clusters, M32 and $\omega$Cen in Fig. \ref{reffmass}. The advantage of the CGS with respect to the simple dissipation-less collapse is that it provides slope and zero point of the observational MRR  and also predicts its change in slope at increasing star mass of the galaxy.

Given these premises, we suggest that the observational MRR, eqn. (\ref{mr1}),  represents the locus on the MR-plane  of galaxies whose formation and evolution closely followed the scheme of dissipation-less collapse, i.e. the ones of the largest mass for each formation redshift. The dwarf galaxies have a different interpretation, because they significantly depart from the above evolutionary scheme and the MRR holding for ETGs. As already pointed out they follow a much flatter relationship which mimics the theoretical MRR (slope 1/3) even if this may be a mere coincidence.

%%%%%%%%%%%%%%%%%%Table 3
\begin{table*}
\begin{center}
\caption{ Coefficients of the polynomial interpolation of the relation (\ref{Lukic_interp}),
which provides the number density of haloes $n(M_{DM}, z)$ per (Mpc/h)$^3$.}
\begin{tabular}{|c|r|r|r|r|r|}
\hline
Mass $[M_\odot/h]$  &     A$_4$      &    A$_3$       &    A$_2$       &    A$_1$    &    A$_0$ \\
\hline
 5e7  &-2.34275e-5  & 	1.28686e-3	  &	-2.97961e-2	  &   2.11295e-1   &  2.02908  \\
 5e8  &-2.76999e-5  & 	1.49291e-3	  &	-3.47013e-2   &   2.13274e-1   &  1.13553  \\
 5e9  &-1.31118e-5  & 	6.50876e-4	  &	-2.36972e-2   &   1.31993e-1   &  0.23807  \\
 5e10 &-1.18729e-5  & 	6.65488e-4	  &	-3.17079e-2	  &   1.30360e-1   & -0.59744  \\
 5e11 &-1.47246e-5  & 	8.10097e-4	  &	-4.65279e-2	  &   1.13790e-1   & -1.44571  \\
 5e12 & 6.59657e-5  &  -7.19134e-4    & -6.99445e-2	  &   1.06782e-1   & -2.45684  \\
 5e13 &-7.34568e-4  &   9.99022e-3    & -1.65888e-1   &  -9.48292e-2   & -3.11701  \\
 5e14 & 4.89975e-3  &  -5.17004e-2    & -1.61508e-1   &  -5.83065e-1   & -4.28270  \\
\hline
\end{tabular}
\end{center}
\label{coef_lukic}
\end{table*}

\section{Simulations of the MR-plane}\label{simul_MR}

The CGS suggests that a deep relation exists between the  way galaxy populate the MR-plane and the cosmological growth of DM haloes and furthermore it  accounts for the slope of the observational MRR. However, it does not provide an explanation for the  tightness of the MRR.   To cast light on this issue, we  analyze the whole MR-plane, applying the same line of reasoning used above, and present a simulation of the MR-plane and the ``observational'' MRR   derived from \textit{first principles}. Once again, we start  from the study of \citet{Lukic2007} and fit their $n(M_{DM},z)$ curves with fourth order polynomial expressions\footnote{Indeed this polynomial fit has already been used to reproduce the \citet{Lukic2007} plane shown in Fig. \ref{lukic}.

\begin{equation}
n(M_{DM}, z) = \sum_{n=0}^4 A_n(M_{DM}) \times z^n \label{Lukic_interp}.
\end{equation}

\noindent The coefficients $A_n(M_{DM})$ are listed in Table \ref{coef_lukic}.}. Then we  count the total number of haloes per mass-bin $\Delta \log M_{DM}$ at redshift $z=0$. This is simply given   by reading off the values of the curves along the $y$-axis and interpolating for intermediate values (we take equally spaced logarithmic bins $\Delta \log M = 0.05$). These are the  haloes that would nowadays populate  the synthetic MR-plane and that should be  compared with the observed galaxies. Of course, to make a meaningful comparison, as the total number of haloes read off the \citet{Lukic2007} plot refers to a volume of 1 (Mpc/$h)^3$, one has to  scale it by  a suitable   factor $C$ to match the real volume of the observed portion of the sky from which the data are obtained. We estimate that $C=5\times 10^6$ is a reasonable choice. This is based on the following arguments: the observational sample we use contains about $\sim 60,000$ objects, i.e.  $6 \%$ of the total number of objects observed by  SDSS ($\sim$ one million); therefore, we should consider $\sim 5-10 \%$ of the total number of haloes expected in the total volume of the sky observed in the survey, which turns out to be $\sim 10^8$ Mpc$^3$, so $C \simeq 0.05 \times 10^8$.  It is worth recalling here that procedure is the same as the one we have been using  to determine the statistics $N_s=10^{-2}$  generating the CGS corresponding the observational MRR. Once more, the line of reasoning we follow may seem  very crude, but it is fully  adequate to the purposes.

Now we have to assign each halo of mass $M_{DM}$ a radius $R_{1/2}$ using eqn. (\ref{mr3}). The radius is also function of the formation redshift $z_{f}$. Therefore, we must assign a typical formation redshift to each halo of mass $M_{DM}$ to be able to plot the corresponding visible galaxy on the MR-plane. To this aim, we proceed as follows. First, we start from the growth functions of \citet{Lukic2007} and derive the number of $n(M_{DM},z)$ of haloes existing in each mass bin $\Delta \log M_{DM}$ at any redshift $z$. This number is the result of two competing effects: the formation of new haloes of mass $M_{DM}$ via merger and/or acquisition of lower mass haloes, and the destruction of haloes of mass $M_{DM}$ because of they merge to form higher mass haloes. Therefore, the following equation can be written

\begin{flalign}
n(M_{DM},z) =& n(M_{DM},z+\Delta z) + \nonumber \\
& n_{+}(M_{DM},z) - n_{-}(M_{DM},z+\Delta z)
\end{flalign}

\noindent where $n_{+}$ and $n_{-}$ represent the creation/destruction mechanisms. In particular, the quantity we are interested in is $n_{+}(M_{DM},z)$, which is the number of new haloes of mass $M_{DM}$ which are born at redshift $z$.

The number of haloes that merge to form higher mass systems is in turn a fraction of the number of haloes existing at that time, i.e. $n_{-}(M_{DM},z+\Delta z) = \eta \times n(M_{DM},z+\Delta z)$, with $0 < \eta < 1$; so

\begin{flalign}
n(M_{DM},z) =& n(M_{DM},z+\Delta z) + \nonumber \\
& n_{+}(M_{DM},z) - \eta \times n(M_{DM},z+\Delta z)
\end{flalign}

\noindent and

\begin{flalign}
n_{+}(M_{DM},z) = & n(M_{DM},z) - \nonumber \\
&  (1-\eta) \times n(M_{DM},z+\Delta z)
\end{flalign}

The only free parameter here is $\eta$, the fraction of haloes that merge to form higher mass systems in the redshift interval $\Delta z=0.145 \times z +0.1$ \footnote{The mass and redshift intervals are as in \citet{Lukic2007}. The expression for $\Delta z$ secures that nearly equally spaced intervals are used for $\log R$.}. In principle, the fraction $\eta$   could vary  with the redshift. However, for the sake of simplicity we assume that $\eta$ remains constant. Thus, we obtain  a value $N_{+}(M_{DM},z)$ for each interval $M_{DM}, M_{DM}+ \Delta M_{DM}$ and $z, z + \Delta z$. This number, re-normalized to  unit over the whole interval, can be considered as the relative probability that a halo of mass $M_{DM}$ is born at redshift $z$.

Finally, for each halo of mass $M_{DM}$ we compare a randomly chosen number $q \in [0,1]$ with the cumulative probability

$$P_{z_i}=\sum_{z=z_{max}}^{z=z_i} n_+(M_{DM},z)$$

\noindent until we have $q<P_{z_i}$, and take $z_{f}=z_i$ as its formation redshift\footnote{The upper limit for the redshift is not always $z_{max}=20$, but the value given by the relation

\begin{equation}
z_{max}=\mbox{INT}(0.0126\tilde{M}^3-0.7597\tilde{M}^2+3.5848\tilde{M}+17.571)
\end{equation}
\noindent with $\tilde{M_z}=M_l-(M_{min}-1)$ obtained by fitting the
\citet{Lukic2007} data.}.
According to eqn. (\ref{mr3}), once $m$ and $f_\sigma$ are known (we adopt here 10 and 1, respectively), the radius of the baryonic  galaxy filiated by this halo,  $R_{1/2}(M_{DM},z_{f})$, is known and the galaxy can be plotted on the MR-plane.

Given these premises, we derive the galaxy counts to be compared with  with the observational data. To this aim,  we divide the MR-plane in a large number of elemental cells with dimensions $\Delta \log M_{s}=\Delta \log R_{1/2}=0.05$, assume a value for the free parameter $\eta$, calculate the number of galaxies expected to fall in each cell, and finally compare this  with the observational counterpart (our SDSS sample).

Before looking at the results, there is an important issue to clarify: the distribution of galaxies on the MR-plane is found to heavily depend on the fraction $\eta$ of haloes which merge to form higher mass systems in the redshift interval $\Delta z$. The point has been checked by adopting different values of $\eta$ and comparing the results with the observational data. We find that in general \textit{the lower the value of $\eta$, the better the simulations agree with the observational data}. For instance, with the present spacing of the  redshift,  $\eta=0.01$ yields excellent results. The agreement would be even better for $\eta=10^{-4}$. In contrast for larger values of $\eta$, e.g. $\eta=0.05$ ($5\%$ of the haloes in each mass bin merge during each redshift interval) the results  significantly worsen. In the following, we will limit to the case with $\eta=0.01$ ($1\%$ of the haloes merge during each redshift interval $\Delta z$), whereas the case with $\eta=0.05$ and an attempt to explain this trend is postponed to Section \ref{mergfreq} below.
The case $\eta=0.01$ can be considered as a sort of acceptable upper limit for this parameter.

\begin{figure}
\centering
\includegraphics[width=.45\textwidth]{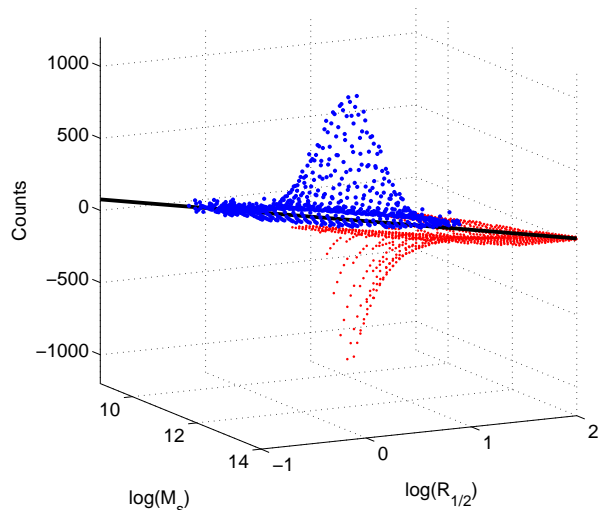}
 \caption{Comparison of the theoretical (red dots, negative  z-axis) with the observational (blue dots, positive z-axis) number frequency distributions of galaxies in the MR-plane: $\log M_S$ is the total stellar mass (in solar units) and  $\log(R_{1/2})$ the half-mass radius (in kpc), ``Counts'' is the quantity ${\Delta n}(z)= {n}(z+\Delta z)-{n}(z)$, i.e. the number of haloes per Mpc$^3$ that are born between $z$ and $z+\Delta z$ plotted per each elemental cell of the MR-plane. The thick line  is the best fit of the observational data:
$logR_{1/2}=0.54\,\,logM_{S} - 5.25$. The observational data are from the HB sample of \citet{Bernardi2010}.
The theoretical number frequencies have been plotted as negative quantities to get the mirror image of the observational data.
} \label{grid_mr_theo}
\end{figure}

The results for the case $\eta=0.01$ are plotted in the 3D space $\log M_s$, $\log R_{1/2}$, $n(M_s, R_{1/2})$ (otherwise referred to as ``Counts'') of Fig. \ref{grid_mr_theo}, in which the previous variables $M_{DM}$ and $z$ are now replaced by $M_s$ and $R_{1/2}$. Along the vertical axis,  the positive values (blue points) show the observational data, whereas the  negative values show the theoretical counts (red points). In this way,   the one-to-one comparison is straightforward. The view angle is chosen in such a way  that the peak values, fall-off toward the high mass range, and thickness of the MRR are simultaneously visible.

%%%%%%%%%%%%%%%%%%%%%%Fig    7

%The theoretical simulation is imposed to fit the peak and the fall-off and slope of the observational MRR towards the high mass end.
We do not make use of the low mass end of the MRR because the observational one is obviously affected by incompleteness and other selection effects. Indeed, while the theoretical expectation is that the number of low mass haloes (low mass visible galaxies) should increase toward the low mass end, the opposite is seen in the data. This difference is most likely due to observational incompleteness even though other  physical causes inhibiting either the formation of stars in low mass haloes (thus making the galaxy hardly observable) or even destroying part of the low mass haloes cannot be excluded. Related to this, the discrepancy between predicted and observed numbers of dwarf galaxies has long been debated with no firm conclusions.  Current hypotheses suggest that small haloes do exist, but fail in producing stars and luminous matter. See below for more details on this issue.

Furthermore, we note that the high-mass tail of the theoretical MRR is too populated with respect to the  observational data. This is likely to be a consequence of the assumption that one halo always hosts one and only one galaxy.
In the real Universe, massive haloes may actually host many galaxies, because merging haloes does not imply merging of their galaxy content. For example, galaxy clusters (of typical masses $M_{cluster}\simeq 10^{14}-10^{15} M_{\odot}$) are known to host hundreds or thousands of galaxies; groups of galaxies can host tens of galaxies, some of which may share a unique dark halo. Thus, in the high-mass tail of our simulation large galaxies corresponding to massive haloes may be replaced by smaller, less massive galaxies, which would  reconcile theory with data. No attempt is taken to cure the theoretical MRR. In any case, the similarity between real and simulated counts is striking.

Looking now at the the thickness of the theoretical and observational MRRs,  at any given radius, while on the high mass side the MRR is bounded by the CGS, its low mass side is determined by the fact that at each halo mass there is a certain redshift below which the number density $n(M_{DM},z)$ starts decreasing: low mass haloes decrease because they either merge into bigger objets or are captured by more massive haloes. As a consequence,  the MRR is expected to be a strip on the MR-plane, populated by galaxies of different type, among which toward the high mass end ETGs (whose structure and evolution closely resembles the dissipation-less collapse) are more numerous. This is clearly shown by the simulation.

\begin{figure}
\centering
\includegraphics[width=9.0cm, height=8.0cm]{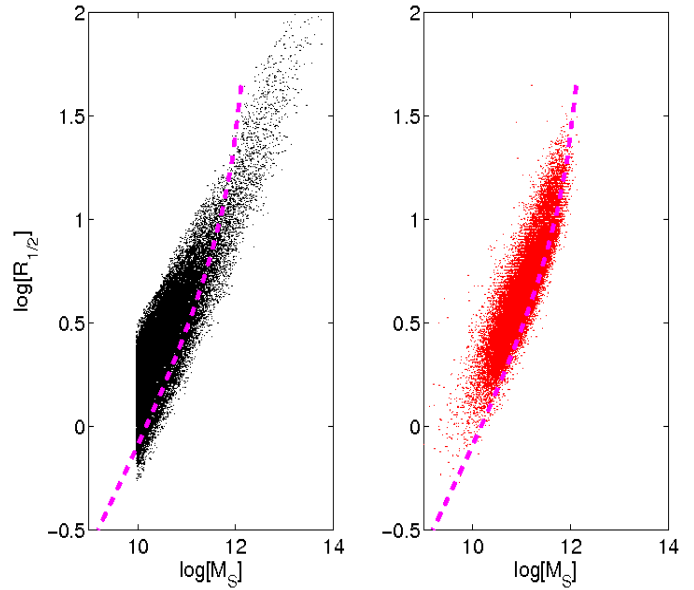}
\caption{Left panel: Predicted MRR limited to galaxies more massive
than $10^{10}\, M_\odot$ for the case with $\eta=0.01$. Right panel: The observational data
of \citet{Bernardi2010} } \label{comp_data_sim}
\end{figure}

After this analysis of the  whole $\log[M_s]$, $\log[R_{1/2}]$, $N(M_s, R_{1/2})$ space, we simulate the MRR of galaxies by drawing as many points in each cell of the MR-plane as indicated by the Counts displayed in Fig. \ref{grid_mr_theo}; we blur the results by artificially applying a small random displacement to each point, considering possible small variations in the $m$ and $f_\sigma$ parameters. The results are plotted in Fig. \ref{comp_data_sim} (left panel), and compared to the real MR-plane of the \citet{Bernardi2010} sample (right panel). Again, the agreement is good; first of all, the simulated points fall exactly in the same region in which real galaxies are found; second, the shape of the two distributions is similar; third, the two distributions have similar width.

As expected, the theoretical distribution is broader than the observational one. The cause is  merely of statistical nature. Indeed there is still something missing so that  the simulated MRR might be  the real counterpart of the observational data. In brief, the previous analysis has estimated the probability that a cell of the MR-plane corresponds to a galaxy with that mass and that radius. To get a realistic simulation of the MRR one has to translate  the  occupation probability of a cell into a number of real galaxies in the same cell. This is particularly true for those cells in which the probability is very low. While on the theoretical side these cells may have a small yet finite occupation probability, in reality with finite numbers of galaxies these cells may be empty. So in reality the MRR is expected to be thinner than predicted by theory. In order to take this into account, we need Monte-Carlo simulations of the MRR. This will be the subject of Section \ref{montecarlo} below.

Finally, the simulated MRR extends towards the realm of the low mass objects and does not find an observational counterpart in the data to disposal. This is less of a problem as the observational sample by \citet{Bernardi2010} contains only objects brighter than a threshold magnitude, so that low mass (and hence low luminosity) galaxies are simply missing. Indeed we have data for dwarf galaxies only in the Local Universe, because observing dwarf galaxies in remote regions of the Universe is still out of reach. Thus, we may argue that many low mass objects should exist, but they are not detected, and hence they are not present in the observed sample. Moreover, we must also keep in mind that our simulations of the MRR stand on the assumption that each DM halo hosts a luminous baryonic object. While for intermediate and high (but not \textit{too} high!) masses this sounds reasonable, at the scale of low mass small galaxies the assumption may brake down. Many low mass haloes could indeed not be able to form visible galaxies, for
instance if the primordial gas cannot collapse into the potential wells of these haloes, because after re-ionization the gas pressure remained high. In other cases, only diffuse, low surface brightness stellar systems may form. Finally, a number of physical processes may concur to dislocate a low mass galaxy from region of the MR-plane in which it was born to other regions of it: for instance, galactic winds, gas stripping, and others (see below).

However, we point out again that the number of small objects which end up populating the simulated MR-plane is consistent with the expectations from a hierarchical theoretical cosmology; indeed, considering the ratios in the number of haloes of different masses at $z=0$ predicted by  the \citet{Warren2006}  mass function, the agreement with our simulation is excellent. Moreover, the same ratios are consistent with those between the number of luminous objects in the Local Group, considering the two massive galaxies, the dwarfs, and estimating the number of Globular Clusters.

%%%%%%%%%%%%%%%%%%%%%%Fig   8

\subsection{Monte-Carlo make-up of the MRR} \label{montecarlo}

To get a simulation as close to reality as possible, we make use of the Monte-Carlo technique and consider $n_{+}(M_{DM},z)$ as the probability that an halo (and the associated baryonic galaxy) with the chosen mass and corresponding radius actually comes to life.

\begin{figure}
\centering
\includegraphics[width=9.0cm, height=8.0cm]{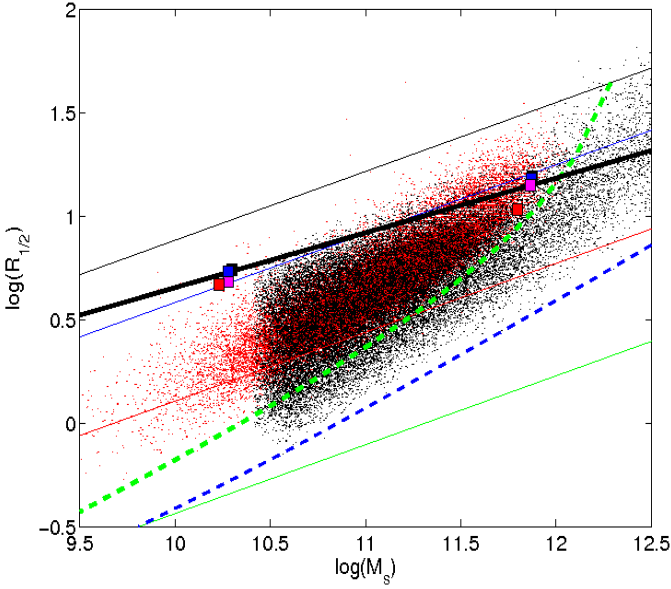}
\caption{Monte-Carlo simulation of the MRR: comparison of the theoretical (black dots) with the observational (red dots) distributions of galaxies in the MR-plane $\log(R_{1/2})$ in kpc versus $\log M_{s}$ in units of $M_{\odot}$. The theoretical simulation is for $\eta=0.01$ (see the text for details). Superposed to it are  the reference models (filled squares) for intermediate and high mass galaxies and their linear fit, the CGS for the  $10^{-2}$ objects per $(Mpc/h)^3$  "statistics":  the thick dashed green line for the stellar component and the blue lines for the Dark Matter halo. The thin trasverse lines are the \citet{Fan2010} relationships  for different redshift, namely z=0, 5, 10, and 20 from the top to the bottom. Finally,    the observational data  are  the HB sample of \citet{Bernardi2010}. } \label{theory_and_data}
\end{figure}

To this aim, we pick a set of $S$ random numbers $r \in [0,1]$, and compare each of them with the value $P = C \times n_{+}(M_{DM},z)$ relative to a randomly chosen cell in the MR plane ($C$ being a suitable normalization factor; the cells are defined as described in Sec. \ref{simul_MR}). If $r<P$, a point is plotted on the plane at the position of the relative cell (with a small random displacement to blur the result), otherwise it is not. The result, adopting $S=3\times10^5$ and $C=10^{-4}$, is plotted in Fig. \ref{theory_and_data}, together with the usual observational sample and with a few theoretical relations.

\begin{figure}
\centering
\includegraphics[width=9.0cm, height=8.5cm]{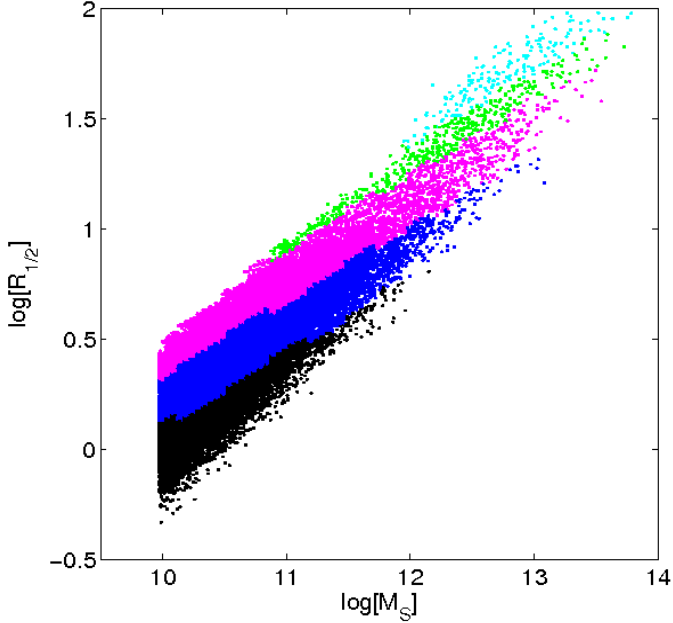}
\caption{The theoretical MRR, $\log(R_{1/2})$ in kpc versus $\log M_{s}$ in $M_{\odot}$; the color code
corresponds to different formation redshifts: cyan $z\leq 0.5$, green $0.5 < z \leq 1$, magenta $1 < z \leq 2.5$,  blue $2.5 < z \leq 5$, and black $z > 5$.  The mean age of galaxies increases from the top to the bottom.
} \label{part_mr_zeta}
\end{figure}

According to our simulations of the MRR, there should be galaxies born at different redshift and hence objects of different age. Fig. \ref{part_mr_zeta} we group the galaxies according to their formation redshift. The mean age in each strip increases moving from the upper to the lower boundary of the MRR. This agrees with the trend found by \citet{Valentinuzzi_etal_2010a,Valentinuzzi_etal_2010b} in galaxies of the WING survey of X-selected galaxy clusters.

%%%%%%%%%%%%%%%%%%%%%%Fig 9

%%%%%%%%%%%%%%%%%%%%%%Fig  10

%%%%%%%%%%%%%%%%%%%%%%Fig  11
\begin{figure}
\centering
\includegraphics[width=9.0cm, height=8.5cm]{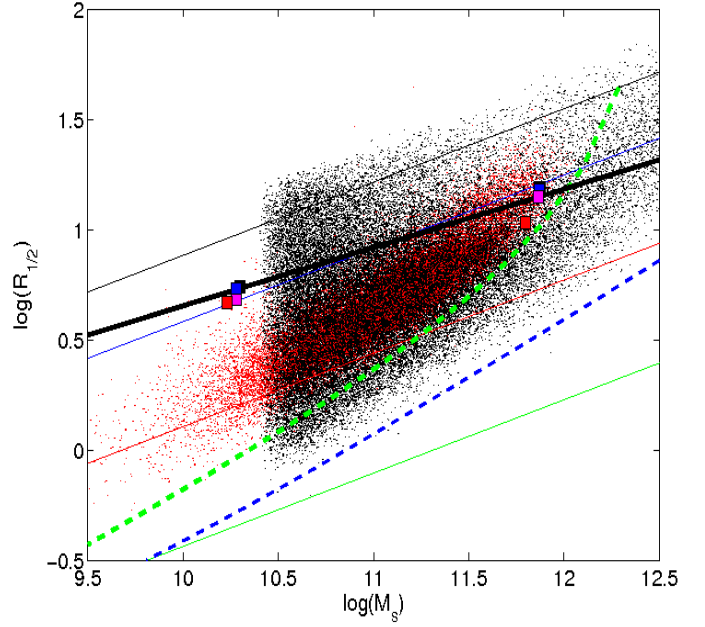}
\caption{The same as in Fig. \ref{theory_and_data} but for $\eta=0.05$. } \label{merg_freq}
\end{figure}

\subsection{Merger percentage}\label{mergfreq}

The agreement we have reached so far has been possible using $\eta=0.01$. As already anticipated, while there is no problem with smaller values of $\eta$, severe difficulties arise increasing it above the value we have adopted. For the sake of argument we show the results for $\eta=0.05$. %(5\% of the present day galaxies can undergo a merger) displayed  in Fig. \ref{merg_freq}.
The least we can say is that theory and data simply disagree. The qualitative explanation of this counter-intuitive result is as follows. Increasing $\eta$ (while keeping constant the total number of haloes of given mass $M_{DM}$ and redshift $z$) induces an increase of the number of haloes of the same mass $M_{DM}$ that collapse exactly at the same redshift $z$; in other words, if $\eta$ is high many haloes merge to form even more massive haloes. Therefore, most of the haloes of mass $M_{DM}$ must be newly collapsed structures. On the contrary, low values of $\eta$ imply that fewer haloes disappear because of mergers, so that a larger fraction of the haloes in place at redshift $z$ must have formed at earlier times and must have survived up to the epoch $z$. As a consequence of this complex game, the distribution function at the epoch of formation of halos with mass $M_{DM}$ changes shape, favouring more recent formation times as $\eta$ increases. In turn, recent formation times imply larger radii (eqn. \ref{mr3}), thus displacing the visible galaxies toward the upper regions of the MR-plane. Basing on these results (excellent agreement between theory and observational data only for rather small values of $\eta$), we are tempted to conclude that mergers cannot have plaid a crucial role in the galaxy formation mechanism.

To cast light on this issue, we assume that $\eta$ is constant with time and calculate the ratio between the total number of merger events, $N_{merg}=\sum_{z_{bin}} \eta \times n(M_{DM},z)$, and the total number of galaxies that ever existed in the Universe over the Hubble time, $N_{tot}=\sum_{z_{bin}} n_+(M_{DM},z)$,

\begin{equation}
{N_{merg} \over N_{tot} } = {\sum_{z_{bin}} \eta \times n(M_{DM},z) \over  \sum_{z_{bin}} n_+(M_{DM},z) }
\end{equation}

\noindent Crude calculations using the \citet{Lukic2007} curves yield $N_{merg}/N_{tot} \simeq 0.10 - 0.15$ for $\eta = 0.01$. This percentage increases to about 50\% for $\eta=0.05$. With the latter value, the price to pay is that the corresponding MRR is much more dispersed than the observational one.  Therefore, our analysis seems to suggest that  only a modest  fraction of haloes should have merged  during the whole history of the Universe.  Owing to the many uncertainties still affecting the above discussion, we do not insist on this  issue. However, it is an important result of this study deserving a thorough investigation.

Different conclusions are reached by \citet{Fakhouri2010} who, using the joint data from the Millennium and Millennium-II simulations, reconstruct the past merger histories of haloes of different mass and estimate their merger frequencies. For halo masses in the range $10^{10}\, M_{\odot}$ to $10^{15}\, M_{\odot}$, ratio of progenitor mass in the range $10^{-5} \leq \xi \leq 1$, and redshift interval $0 \leq z \leq 15$, they find the merger rate per unit redshift to be nearly independent of $z$ and halo mass, and to depend on $\xi$ as $\propto \xi^{-2}$. Finally, referring to a typical halo mass of $10^{12}\, M_{odot}$ today, they estimate the probability of a major merger since $z$=1, 2 and 3 to be about 31\%, 53\% and 69\%, respectively. Haloes with the same mass at $z$=2, have about 50\% probability of major mergers since $z$=4. %What we learn from \citet{Fakhouri2010} is that the probability per halo and unit redshift is nearly constant, which could be equivalent to assuming $\eta$ to be constant in our formalism. 
The above probabilities refer to a specific halo mass, so the translation to the percentage of mergers with respect to the whole galaxy content over the entire history of the Universe is not straightforward.

% merger frequencies of about 50\% and higher. 
The only remark we can make, is that perhaps the two estimates could indeed be closer than it might seem at a first sight. The merit of our estimate is that observational data are used to compare the model results (choice of $\eta$ constrained on the MRR), whereas the \citet{Fakhouri2010} analysis and estimates of the merger percentage are theoretical, with little reference to the real world of galaxies we see in the Universe. Surely the topic is very alive and worth being pursued.

\subsection{Possible effects blurring the MRR}

We have shown that a tight MRR may result from first physical principles coupled to elementary cosmology. For the sake of completeness, we may turn the argument around and check whether starting from an intrinsically sparse distribution of galaxies on the MR-plane, one may recover a tight MRR as the result of some physical effects.
In other words, there are at least two important phenomena at work, i.e. mergers of two galaxies of different mass (and size) into a single object and matter ejection by a single galaxy by sudden injection of energy (e.g. AGN activity and/or
galactic winds) that could change the position of a given galaxy on the MR-plane and therefore scatter the expected distribution. Whether, on the contrary, starting from a scattered distribution of galaxies, they can give rise to more or less ordered sequence such as that indicated by the MRR is hard to prove, and unlikely to occur in the real world.

\textsf{Minor and major mergers}. In the hierarchical scenario, minor (not equal masses) and major (equal masses) mergers of low mass galaxies into ones of larger masses is the paradigm mechanism by which galaxies acquire dimensions and masses that we observe. There are many papers and even text books trying to resolve this issue, e.g. \citet[][]{Binney_Tremaine2008,Hopkins_etal_2008,Naab_etal_2009} to mention a few. The formalism in use here is taken from Bernardi (2011, talk delivered at the University of Padova, private communication). The process can be accompanied by star formation (so-called \textit{wet merger}) which would spread the age and the metallicity of the stellar populations hosted by galaxies. This may cause some difficulties with the ETGs whose stars seem to be old with a rather narrow age range. We will come back to this later on. In alternative, the mergers can occur free of star formations (in this case they are referred to as \textit{dry mergers}). Wet or dry, by how much a merger between two bodies of masses $M_1$ and $M_2$ and radii $R_1$ and $R_2$ can increase the mass and size of the daughter galaxy? For a pair of
interacting virialized galaxies made of DM, we may assume that the total energy is given by the sum of kinetic and gravitational potential

$$E=E_{vir} + E_{int} \simeq E_{vir} = K_{vir} + W_{vir}$$

\noindent where $E_{vir}$ is the total virial energy of the two galaxies and $E_{int}$ is the energy of interaction. This can be neglected with respect to the first one so that the total energies before merging is

$$ E_i = {M_{1}\sigma_{1}^{2}  \over 2 } + {M_{2} \sigma_{2}^{2} \over 2}
- { G M_{1}^{2} \over R_{1} }  - {G M_{2}^{2} \over R_{2} }$$

\noindent where the $\sigma$'s are the velocity dispersions. After merging the total energy of the system is

$$ E_f = {(M_{1} + M_{2})\sigma_{f}^{2} \over 2}   - {G (M_1+M_2)^{2} \over R_{f} } $$

\noindent In a dissipation-less process, during the merger the total energy is conserved, so $E_f=E_i$. In the case of a major merger represented by $M_1 = M_2=M_i$ and $M_f=2M_i$ we get

 $$\sigma_i^2  -   {G (2 M_i) \over R_i} = \sigma_{f}^{2}  -  {G (2 M_f) \over
 R_f} $$
which means that the mass and radius are doubled, whereas $\sigma$ remains unchanged. In the MR-plane the displacement vector has slope $\Delta \log R /\Delta \log M =1$ \citep{vanDokkum2010}.

\noindent In the case of a minor merger represented by $M_f = (1+f) M_i$, where $f$ is smaller than 1, the  Virial Theorem ($2K = -W$ with obvious meaning of the symbols) yields $M_i ~ R_i \sigma_i^2$, from which we get

$$R_f \sigma_f^2 = (1+f) R_i \sigma_i^2 = (1+f)^2 R_i \sigma_i^2 /
(1+f)$$

\noindent When $f << 1$  we obtain

$$M_f = (1+f) M_i \simeq (1+2f) R_i \sigma_i^2 (1-f)$$

\noindent i.e. the change in size is larger than the change in mass and also $\sigma$ decreases. The displacement vector has slope $\Delta \log R /\Delta \log M > 1$ likely close to 2 \citep{vanDokkum2010}.

There is a final remark worth being made. The above formalism has been developed looking at the sole DM component
of the interacting galaxies. How the BM component will behave during the merger event neither is completely understood nor enough simulations are available to make solid predictions. Most likely long trails of DM and BM will develop, thus likely removing significant amounts of mass of the two components. The effect can be particularly relevant for BM because of the many implications for subsequent star formation (if any). In any case, mergers alone cannot be responsible of the well behaved MRR, but would simply add some dispersion on it.

\textsf{Puffing up and galactic winds}. Rapid stripping or ejection of BM from galaxies may puff them to larger dimensions. The idea goes back to \citet{BiermannShapiro1979} to explain the formation of S0 galaxies from disc galaxies. Recently it has been reconsidered by \citet{RagoneGranato_2011} to explain the observational evidence that most massive ETGs at redshift $z\geq 1$ exhibits sizes smaller bay a factor of a few than local galaxies of the same type and  mass. They investigate the effect of BM loss triggered either by QSO/starburst/driven galactic winds or quiet gas
restitution and ejection by stars at the end of their evolution, both concurring in a galaxy to eject a sizeable fraction of the BM mass. The evolution of the galaxy depends on the ejection timescale compared to the dynamical timescale. If the former is short compared to the latter (fast ejection) the ratio of the final to the initial radius is

$$ { R_{f} \over R_{i}} = {(M_f/M_i) \over 2(M_f/M_i) - 1 }$$
and if $M_f/M_i \leq 0.5$ the system gets unbound and dissolves. If the ejection time scale is long then the conservation of adiabatic invariants yields

$$ { R_{f} \over R_{i}} = {1 \over (M_f/M_i) }$$

\noindent Thus, if the ejection is fast the size increase is more efficient, if the ejection is slow the system remains bound independently of $M_f/M_i$. In this scheme the compact objects should transform in less massive and larger systems. The process could occur in two steps: a very fast one at the beginning of the star forming history a slow one during the remaining life. Indeed our NB-TSPH models partly follow this scheme, and partly not. The models have intense episodes of star formation and significant galactic winds but on the average the general trend is toward larger star masses and nearly constant radii. Whether this mechanism can account for the general distribution of galaxies on the MR-plane is hard to say \footnote{The discussion by \citet{Chiosi2002} on this issue is still valid, and it is strengthened by the new NB-TSPH models used in this study. In their analysis, \citet{Chiosi2002} showed that the discrepancy between the slopes of the MRR traced by models and  data for ETGs  cannot be ascribed to mergers, because in this case the models would fall even farther out with respect to existing observational data.}.

\textsf{Geometrical effects}. \citet{Guo2009} and \citet{vanDokkum2010} have investigated another interesting possibility: the MRR, at least for the most massive galaxies, is driven by a systematic variation of the Sersic index $n_S$ parameterizing the surface density profiles with the redshift. According to \citet{vanDokkum2010}, in the context of the inside-out growth of a galaxy by accumulation of material, the variation of the effective radius $R_e$ (50\% of the light) and $R_{1/2}$ in turn, the Sersic index, and the galaxy mass is

$$ {d\log R_{e} \over d\log M } \approx  3.56 \log(n + 3.09) -1.22$$

\noindent accurate to 0.01 dex for $1 \leq n \leq 6$. This means that the radius increases approximately linearly with the mass if the projected density follows an exponential law, but that $\propto M^{1.8}$ for the de Vaucouleurs profile with  $n_S=4$. This implies that strong evolution in the measured $R_e$ is expected in all inside-ut growth scenarios irrespective of the physical mechanism, unless the density profiles are close to exponential. In our models the Sersic index varies from $ n_S\simeq 2$ for the HDMM case to $n_S=4$ for the HDHM one. Therefore the expected variation of the slope ${d\log R_{e} \over d\log M }$ due to  geometrical effects goes from about 1 to about 1.8. This implies that the displacement vector changes slope according to the Sersic index of the underlying galaxy. The slope increases from $R_e \propto M$ to $R_e \propto M^2$.
Indeed  looking at the mean locus of galaxies in the MR-plane, the slope increases moving toward the top right corner of the plane, in other words the band tends to bend toward higher radii as the
mass increases, passing from 0.5 for the low mass galaxies, to 1 for the intermediate mass ones, to about 2 for those at the top end.  \citet{vanDokkum2010} made use of the above relation to explain the steep slope of the MRR for galaxies measured at different redshifts (from $z=2$ to $z=0$) whose radii and masses seem to increase according to $R_e \propto M^2$: $0.45 \leq \log R_e
\leq 1.1 $ and $11.15 \leq M(M_\odot) \leq 11.45$. At high redshift ($z \simeq 2$) these galaxies have smaller effective radii and profiles that are closer to exponential, whereas at low redshift ($z\simeq 0$) the radii get normal and the profiles approach the de Vaucouleurs law. It seems that the size and hence the density is correlated with the
Sersic profile which in turn drives the displacement vector in the MR-plane. However, it is worth recalling that exactly the same trend of the MRR is also shown by the CGS in the mass range of interest, thus partially weakening  the argument based on geometrical effects. Likely both concur to shape the top end of the MRR.

%%%%%%%%%%%%%%%%%%%%%%%%Fig 12
\begin{figure}
\centering{
\includegraphics[width=9.0cm,height=8.0cm]{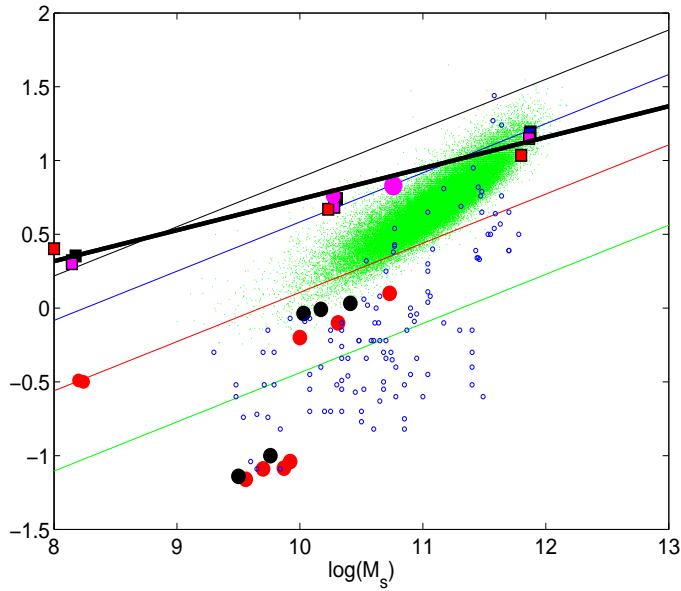}}
\caption{The $\log[R_{1/2}]$ versus $\log[M_{s}]$ relation for compact galaxies (open circles).  The green dots  are the normal ETGs of the HB sample of \citet{Bernardi2010} plotted here for the sake of reference.    The squares are the twelve reference models together with their linear fit (black line), whereas the  filled circles are the ancillary models with different efficiency of  star formation. The thin straight lines are the theoretical MRR on which proto-galaxies of given DM over-density settle down at the collapse stage as a function of the redshift. The lines are given by eqn. (\ref{mr3}) taken from \citet{Fan2010} and correspond to   $m=10$ and $f_\sigma=1$; the color-code indicates  different redshifts (black: $z=0$; blue: $z=1$; red: $z=2.5$; green: $z=5$; cyan: $z=20$). The compact galaxies could correspond to models in which star formation has been  inhibited while the gas was falling into the gravitational potential well of DM+BM toward higher and higher densities. The models are described in Appendix A and Section 3.  }
\label{rm_comp}
\end{figure}

\section{The compact objects at high redshift}\label{compacts}

As already anticipated the existence of compact, massive galaxies is an established fact which seems to be relatively frequent.  Although the detection of compact galaxies at high redshift may be affected by observational biases \citep[e.g.,\,][]{Mancini_etal_2009}, extremely deep imaging \citep[e.g.,][]{Szomoru_etal_2010}, lensing \citep[e.g., ][]{Auger_etal_2011, Newton_etal_2011}, available measurements of very high velocity dispersions
\citep[e.g.,][]{CenarroTrujillo2009,Cappellari_etal_2009,vanDokkum2009,vandeSande2011}, and measurements of the surface brightness \citep{Szomoru2011}, are confirming their compactness. Furthermore, according to \citet{Saracco_etal_2011} compact galaxies tend to complete their star forming activity at $z> 5$, whereas larger galaxies of similar
mass tend to have a younger stellar content suggesting that the structure of a galaxy may be related to their star formation activity \citep[see][ and references therein]{Mosleh_etal_2011,Shankar_etal_2011}.

The sample of compact galaxies we are going to consider is gathered from different sources and spans an ample range of redshifts.  Data for low mass objects   is  taken from  \citet{Cimatti2004}, \citet{Daddi2005}, \citet{Maraston2006}, \citet{Toft2007}, \citet{Zirm2007}, \citet{Cimatti2008},
\citet{McGrath2008}, \citet{vanDokkum2009}, \citet[][]{Gadotti_2009}, \citet{Fisher_Drory_2011}, and see also
\citet{Shankar_etal_2011}. The masses (in $M_{\odot}$) and radii (in kpc) fall in the ranges $9 < \log M_s  < 11$ and $0.3 < \log R_{1/2} < 1.3$, respectively. They define a mean relation roughly represented by

$$ \log R_{1/2} = 0.59 \log M_s - 6.51$$
It is worth emphasizing that this relation runs nearly parallel to that drawn by the SDSS data of \citet{Bernardi2010}.

On the high mass range we have data from \citet{Bezanson2009}, \citet{Damjanov2009}, \citet{Mancini2010}, \citet{Longhetti2007}, and recently \citet{Tiret2011}. The estimated masses and radii (in the same units) span the intervals $11 < \log M_s  < 12$ and $0.4 < \log R_{1/2}  <  1.3$. They define a mean relation given by

$$ \log R_{1/2} = 1.07 \log M_s -11.95$$

\noindent Remarkably the slope now is nearly twice as much as that in the lower mass interval and intersects the region of the most massive and widest ETGs in the \citet{Bernardi2010} sample.  For more details on the subject and the current explanation for both compact and extended objects at the same redshift (similar luminosity and hence mass) see \citet[][]{Shankar_etal_2011}.

The position of the compact galaxies on the MR-plane  is shown  Fig. \ref{rm_comp} together with normal ETGs. Even if a formal comparison between the two samples cannot be made because the data are no fully homogeneous, yet an idea of their relative position can be inferred.  Concerning compact galaxies, there are at least three questions to be answered: How can these compact (and high mass) objects exist at large redshift? How do they evolve? Which are their counterparts in the local Universe?

The position of these objects on the MR-plane and comparison with the iso-density lines would imply a rather high redshift of formation to which a higher initial density corresponds. The upper limit is of the order of $z\simeq 20-25$. Therefore, there are two issues to address: can galaxies with the mass currently assigned to compact galaxies exist at these redshifts? Is their position indicating the initial density or what else?

Compact galaxies with estimated star mass in the range $ 10^{10}$ to $5\times 10^{12}\, M_\odot$ --  see for instance
the data by \citet[][]{Gadotti_2009}, \citet{Bezanson2009}, \citet{Damjanov2009}, \citet{Mancini2010}, \citet{Fisher_Drory_2011}, and recently  \citet{Tiret2011} -- likely  belong to systems made of DM + BM of significantly bigger mass that can be estimated from $M_{DM} =m\, M_s$ with $m=15 \div 25$. We adopt here $m=10$
over the mass range $10^{10} \leq M_s \leq  10^{13}$. According to \citet{Lukic2007}, the probability of occurrence per (Mpc$/h)^3$ for galaxies with total mass in the range $(10^{10}-10^{11})M_\odot$ goes from a few $10^{-1}$ at $z\simeq 1$ to below $10^{-5}$ at $z\simeq 16$, whereas for objects in the mass range $(10^{12}-10^{13})M_\odot$ the probability falls from a few $10^{-3}$ at $z\simeq 1$ to below $10^{-5}$ at redshift $z\simeq 11$. These probabilities albeit small are still significant for a total observed volume of about $10^8$ Mpc$^3$.

Similar considerations can be made for the objects in the lower range of mass even though they have a larger probability of coming into existence at high redshifts. So massive galaxies with very high initial density (high formation redshift) are possible.

Looking at the position of the ancillary models on the MR-plane of Fig. \ref{rm_comp}, two possible explanations can be put forward: i) very high redshift and hence very high mean initial density; (ii)  more normal redshift and hence initial over-densities but intrinsically less efficient star formation. This is what indeed is shown by our ancillary models which show that in both cases more compact objects are to be expected. The case of high redshift, high initial densities is in a sense following the general rule. It could correspond to the case generating globular clusters and/or bulges of galaxies.  On the contrary,  in the current view of galaxy formation, the role played by the star formation threshold density has never been fully highlighted and therefore deserves some attention.    If for any
reason star formation is inhibited at the gas densities that were good for normal galaxies, the primordial gas can collapse in the potential well of DM+BM to higher densities and smaller radii in turn. When star formation begins the dimensions of the galaxy get frozen according to our simulations. If the sequence of compact galaxies is generated when a new density threshold is reached by the BM inside a DM halo, we expect that it should reflect the mean behaviour of the sequence of normal galaxies. We have already argued that this latter mirrors the shape of the CGS on the MR-plane. Indeed the sequence of compact galaxies changes its slope from 0.5 to 1 and even higher as the mass goes toward the upper limit. Mergers and mass ejection episodes may occur thus inflating the system as already reported above and therefore blur the ideal sequence.  This explanation in terms of the threshold density for star formation does not contradict the suggestion that these objects could be the bulges of bigger galaxies \citep{Graham2011}.
Finally, there is another parameter to consider, i.e. $f_\sigma$. According to eqn. (\ref{mr3}), at given total mass, redshift and all remaining parameters, the $R_{1/2}$ radius scales with $(1.5/f_\sigma)^2$. Passing from the reference case $f_\sigma$=1 to $f_\sigma$=2, the radius goes down by a factor 0.5. Therefore, given the position of a compact object on the MR-plane the associated initial redshift would turn out to be about a factor of two  smaller than before, thus alleviating the initial redshift required to explain the position of compact galaxies on the MR-plane.
The subject is left to future investigation.

\section{Summary and conclusions} \label{conclusions}

Before drawing some general conclusions about the results we  have obtained, a few remarks are appropriate. First of all, our NB-TSPH galaxy models can be classified as \textit{early hierarchical or quasi monolithic} because only during the very early stages there exists a hierarchy of substructures merging together into a single object. The formation of a massive galaxy is completed within the first two Gyr or so \citep[see][for more details]{Merlin2012}. Second,  the recent discoveries of high mass galaxies ($M_s\simeq10^{11} M_{\odot}$) at high redshift ($z \simeq 7-8$) do not contradict theoretical predictions in which the existence of massive galaxies is possible even at high redshifts. Indeed, looking at the MR-plane (Fig. \ref{theory_and_data}), the iso-density line corresponding to the collapse redshift $z=5$ intersects the CGS at a mass $M_s\simeq 5\times10^{11} M_{\odot}$, so that there is ample room for high mass galaxies. Third, the galaxy models in use clearly follow the so-called downsizing pattern, originally predicted by \citet{Cowie1996},  later confirmed by the NB-TSPH models of \citet{Chiosi2002}, and more recently amply supported by a plenty of  observational data  \citep[see][for a recent discussion]{Bundy2006}. They find that over the probed redshift range, (i) there is a mass limit above which star formation appears to be quenched; (ii) the mass limit decreases with time by a factor of 3 across the redshift range; (iii) the process for quenching star formation must, primarily, be internally driven. We point out that the arguments given by \citet{Bundy2006} do not contradict the \textit{early hierarchical or quasi monolithic} view in which galaxies evolved quiescently after the initial period of strong mass assembly and star formation. The time interval span by this early activity was indeed rather short, the maximum duration being the time interval elapsed from $t(z_f)$ to $t(z=1.4)$. If they later suffered one or more mergers, most likely these were not accompanied by star formation with no consequences on their stellar content. In other words, an \textit{early-hierarchical} scenario cannot be ruled out and most likely took place indeed. It is wise to say that early-hierarchical, quasi-monolithic and more rarely pure hierarchical schemes all concur to shape the present day population of ETGs; however, the hierarchical scheme cannot be the sole dominant channel of galaxy formation.

In this context, the main results and conclusions of this study are:

\begin{itemize}
\item
The observational distribution of galaxies on the MR-plane does not coincide with locus expected for haloes and companion baryonic galaxies on the basis of their mass and radii at assigned mean density. Rather, we think that it results from two complementary mechanisms:

(i) As the Universe expands, haloes of higher and higher mass collapse.  The radius of an halo is proportional to its mass according to the relation $R_{DM}\propto M_{DM}^{1/3}$ (iso-density line). The proportionality factor depends on the cosmological background and the  virialization redshift because systems that virialize first have larger mean initial densities (the Universe was denser when they collapsed). In other words, the initial density of an halo is defined by the background density at the moment of its virialization. Inside these haloes, the baryonic matter collapses and soon or later is turned into stars at a suitable rate, fixing the ratio $m=M_{DM}/M_{s}$. So there is a manifold of MRRs described by $R_{1/2}$ and $M_s$ of a BM galaxy parameterized by the initial density, i.e. the redshift of the initial collapse. If the formation of a BM galaxy inside a DM halo is nearly dissipation-less, the process occurs at  constant gravitational potential energy per unit mass of the DM and BM components. In general, massive galaxies follow this rule, so that  the parent DM haloes and the daughter BM galaxies lay on nearly parallel  MRRs, whose slope is close to  $1/3$. At decreasing total mass of a galaxy, the dissipative processes become more and more important so that  the MRR slope decreases  from 0.333 for galaxies with total mass $10^{13}\, M_\odot$ to about 0.2 for a total mass of $10^9\, M_\odot$. This latter slope nicely coincides with the observational mean value of the MRR for dwarf galaxies which are more sensitive to the complex physics of baryons (cooling, heating, galactic winds, etc.).

(ii) The cosmology background fixes the ``upper mass'' of the haloes collapsing at any redshift, i.e. the upper mass end of the mass distribution. Assuming a value for the total number density of haloes, we derive from the number density-plane the CGS to be plotted on the MR-plane in coordinates $R_{1/2}$ vs $M_s$. The CGS is the locus of the ``typical masses'', function of the redshift, limiting on the high mass side the manifold of theoretical MRRs. For $10^{-2}$ haloes per (Mpc/$h)^3$ the CGS fits astonishingly well the high mass boundary of the observational MRR and nicely reproduces its slope. Furthermore, the slope of the CGS and that of the observational MRR coincide with the relationship predicted by  the top-hat spherical dissipation-less collapse for primordial fluctuations originally proposed by \citet{Gott_Rees1975} and further developed by \citet{Faber1984} and \citet{Burstein1997}. This cannot be a mere coincidence.

\item The observational MRR and simulations of the MR-plane based on the number densities of haloes $n_+(M_{DM},z)$ are in close agreement. The lower boundary of the MRR strip is set by the ever increasing typical mass of the underlying mass distribution of the collapsing DM haloes, i.e. the CGS. In other words, at any given value of the redshift, i.e. along the corresponding MRR, only values below the typical mass are permitted. Haloes with higher masses are so improbable that the region of the MR-plane below the CGS is simply void of galaxies with exception of the so-called ``compact galaxies'', for which a plausible explanation not in contradiction with the previous one can be found. The upper boundary of the strip is set by the slope of  growth function of a certain halo mass  becoming negative below a certain value of the redshift (haloes of that mass start decreasing) plus other  physical processes of minor importance.

\item Within the mass interval of existence for haloes formed at a given  redshift, those with the typical mass  form a baryonic galaxy inside closely following the ideal case of the dissipation-less collapse. This latter point and the consideration made above eventually determines the zero point and the thickness of the MRR.

\item The low mass end of the simulated MR-plane is over-populated by small objects, with respect to the observational data. As explained above, this is not a problem, and is instead consistent both with theoretical expectations and  a rough counting in the very local Universe (the only region of the cosmos in which we can be confident to really obtain complete catalogues!). On the other hand, the high mass end of the MRR contains galaxies whose total mass  falls in a risky mass range, more typical of the mass of groups or small clusters than of individual objects. Indeed, we do not observe galaxies more massive than say $M_s\simeq 10^{12} M_{\odot}$ (the associated halo mass would be $\sim 10^{13}M_{\odot}$). In contrast we find haloes with masses up $M\simeq10^{14}-10^{15} M_{\odot}$ which in turn are hosting groups and clusters of galaxies. In this case we may invoke the assumption of the one-to-one correspondence between halo mass and baryonic galaxy inside, and argue that haloes with mass above a given threshold value do not contain a single ultra-massive galaxy (with the possible exception of the CD galaxies in clusters for which recent dry mergers are likely responsible)) but rather host a few smaller galaxies. Correcting for this effect  would likely yield the right number of galaxies also in this region of the MR-plane.

\item According to our simulations of the observational MR-plane at each redshift the percentage of galaxies undergoing mergers should be small, otherwise the theoretical MRR would not agree with the observational one. More than this we cannot say with the present analysis. This is an unexpected important result deserving thorough investigation.

\item Finally, we like to mention that the present analysis essentially confirms the suggestion advanced long ago by \citet{Chiosi2002} on the base of a much smaller sample of observational data and much simpler NB-TSPH galaxy models.

\end{itemize}

\textbf{Acknowledgements}
The authors are deeply grateful to Dr. Mariangela Bernardi for providing easy access to the observational data and the many useful clarifications and discussions.
%\end{acknowledgements}

%=================== BIBLIOGRAPHY ======================%

\bibliographystyle{mn2e}           % Files .bst

\bibliography{mnemonic,Mass_Radius_biblio}    % Files .bib

\section*{Appendix A}\label{appendixA}

\section*{NB-TSPH  models of ETGs}

\citet{Merlin2010,Merlin2012} with aid of the  parallel NB-TSPH code EvoL  produced a number of models for ETGs with different mass and/or initial density (twelve cases in total). The models are followed from the epoch of their detachment from the linear regime, i.e. $z_{i} \geq 20$, to a final epoch (redshift $z_f$) varying from model to model (however with $z_{f}\leq 1$). The simulations include radiative cooling down to 10 K (with sets a suitable artificial pressure floor to avoid spurious clumping of particles), star formation, stellar energy feedback modeled with a novel method (the Almost-Zero-Mass Particle method),  re-ionizing photo-heating background, and chemical enrichment of the interstellar medium. The reader can refer to the cited papers for all the details on the method and the code in use as well as on model results.

The assumed cosmology is the standard $\Lambda-$CDM, with $H_0$=70.1 km/s/Mpc, flat geometry, $\Omega_{\Lambda}$=0.721, $\sigma_8$=0.817; the baryonic fraction is $\simeq 0.1656$.

Each NB-TSPH model initially consists of $\simeq 60,000$ DM particles plus an equal number of gas particles, a fraction of which are turned into stars during the simulation. A gas particle is eligible to form stars if its density exceeds a threshold value ($\rho_{sf}=5\times10^{-25}$ g cm$^{-3}$) and its velocity divergence is negative; then, a Monte-Carlo stochastic selection is adopted to decide which particles effectively turn into stars.

Each model starts as a sphere of DM plus gas particles, cut from a wider cosmological simulation in which  density fluctuations exist and the at the center  is a constrained central peak of given density contrast of the cosmological tissue. The central density peak has a given total mass $M_T$, sum of the DM and BM components, ranging from $\sim 10^9$ to $\sim 10^{13} M_{\odot}$. This is the proto-halo of our model galaxy inside which stars will be formed at later times. The cosmological simulation provides the initial positions and velocities of all the particles in the prot-halo.  A minimal amount of rotation  and an outward radial component meant to mimic the expansion of the Universe are added to all the particles.  The initial conditions are set in such a way  that each model is a re-scaled version of a single reference proto-galactic halo, with different total mass and/or initial over-density. In this way, the attention is focused specifically on the role played by different initial masses and densities rather than by local inhomogeneities. The properties of all the twelve haloes are listed in Table \ref{tab1}.

The proto-galactic haloes are then followed through their early stages of expansion following the Hubble flow, the turn around, and the collapse. The redshift at which the collapse begins varies from model to model and inside the same model from the center to the outer regions. In general the collapse occurs during the redshift interval $4 > z > 2 $, it starts first in the central regions and gradually moves outwards. The collapse is complete at redshift $z\simeq  2$.

All the models develop a central stellar system, with a spheroidal shape. They confirm the correlation between the initial properties of the proto-haloes and their star formation histories already found by \citet{Chiosi2002}. Massive haloes ($M_{T}\simeq 10^{13} M_{\odot}$) experience a single, intense burst of star formation (with rates $\geq 10^3 M_{\odot}$/yr) at early epochs, consistently with observations, with a less pronounced dependence on the initial over-density; intermediate mass haloes ($M_{T}\simeq 10^{11} M_{\odot}$) have star formation histories that strongly depend on their initial over-density, i.e. from a single peaked to a long lasting period of period of activity with
strong fluctuations in the rate; finally small mass haloes ($M_{T}\simeq 10^{9} M_{\odot}$) always have fragmented
histories, resulting in multiple stellar populations, due to the so-called ``galactic breathing'' phenomenon.

The models have morphological, structural and photometric properties comparable to real galaxies, in general closely
matching the observed data; there are minor discrepancies that are likely of numerical origin \citep[see][\, for all details]{Merlin2012}.

These models can be classified as \textit{early hierarchical} because they undergo repeated episodes of mass accretion of sub-lumps of matter inside the original density contrast in very early epochs and essentially evolve in isolation ever since. At the present time, the models closely resemble real galaxies. All this leads us to conclude that the evolutionary scheme \textit{early aggregation followed by isolation} (thus much resembling the classical monolithic scheme) is able to explain many of the observed features of ETGs, particularly the complicated and different star formation histories shown by haloes of very different mass, without invoking major, late mergers (i.e. the classical hierarchical scheme).

The reference models are calculated adopting a star formation efficiency $\epsilon_{sf} = 1$. This value is  larger than current estimates from observational data in the local Universe, i.e. $\epsilon_{sf} \simeq 0.025$ \citep{Lada2003, Krumholz2007}, and theoretical considerations on the global star-to-total mass ratio in galaxies suggest $\epsilon_{sf} \leq 0.1$. However, adopting a high value of $\epsilon_{sf}$ allows for a strong reduction of the computational time, while preserving the basic properties of the models \citep[for a complete discussion see][]{Merlin2012}.

For the purposes of this study, using the same numerical code and approach for the initial conditions, we also produce a group of \textsf{ancillary models}, whose initial parameters and key results are listed in Table \ref{tab2}. These models are calculated to explore the consequences of much higher initial density contrasts and/or lower star formation efficiencies $\epsilon_{sf}$.  The the effect of the higher initial density is already known from the old calculation \citet{Chiosi2002} and the models by \citet{Merlin2012}. Galaxies of the same mass will be shifted on the MR-plane to smaller radii. The effect of star formation efficiency is less certain. As expected by changing (decreasing) the efficiency $\epsilon_{sf}$ star formation is delayed or even inhibited. The ga continues to flow into the gravitational potential well till the threshold density for star formation to occur is reached and /or s sufficient number of stars are formed, the newly born galaxy has much smaller dimension with respect to the corresponding object with higher efficiency of star formation.  All the ancillary models  are calculated limited to the  very early evolutionary stages, to show in the MR-plane the initial position of a model galaxy in which the gas content has reached densities much higher than the formal mean background density fixed by cosmology. The parameter  $f_\delta$ indicates the factor by which the initial density of same model in the first group (Table \ref{tab1}) is scaled. Of course by doing this ones has to recompute the initial redshift $z_i$ and all other parameters using the standard procedure based on large scale cosmological simulations. As already said in these models we also change the efficiency of star formation $\epsilon_{sf}$ as indicated in column (3) Table \ref{tab2}. In a few models, indicated by $\epsilon_{sf}(z)$, the efficiency of star formation increases with metallicity $Z$ of the gas content according to
$\epsilon_{sf} = MIN(1.,10.^{(.5\,log(Z)+1.)})$ on the notion that a metal-rich gas finds it easier to collapse and form stars. The efficiency goes from $\epsilon_{sf}=0.1$ for $Z$=0.0001 to $\epsilon_{sf} = 1$ for $Z$=0.01 (close to the solar value).  No special meaning must be given to this relation, it is simply meant to evaluate the effect of an efficiency of star formation increasing with the metallicity. In any case, this effect plays a marginal role on the position of the model galaxies on the MR-plane, see the entries of Table \ref{tab2}. The effects of higher initial density and efficiency of star formation passing from $\epsilon_{sf}=1$ to 0.1 or so are of paramount importance and cannot be ignored.

%%%%%%%%%%%%%%%%%%%%%%%%%%Table 1
\begin{table*}%%%%%%%%%%%%%%%%%[h!]
\caption{The twelve reference models. Left to right: total initial mass $M_T=M_{DM} + M_{BM} $ [in units of $10^{12} M_{\odot}$], corresponding (initial) gas mass $M_g$ [in units of $10^{12} M_{\odot}$], initial redshift $z_i$, mean halo over-densities $[\delta\rho-1]_{i}$ at the initial redshift, initial proper physical radius of the halo [in kpc], redshift of the last computed model $z_f$, corresponding age $t_f$ [in Gyr], virial radius of the whole system $R_{vir}$ [in kpc], half-mass  radius $R_{1/2}$ at $z_f$, and the total stellar mass $M_s$ at final redshift [in units of $10^{12}M_{\odot}$]. } \centering
\begin{tabular}{|l| l|l|l|c|l|l|l|l|l|l|}
\hline
Model & $M_{T}$  & $M_{g,i}$ & $z_{i}$ & $[\delta\rho-1]_{i}$ &  $R_{i}$ & $z_{f}$ & $t_{f}$ & $R_{vir}$ & $R_{1/2}$ & $M_{s}$  \\
\hline
HDHM  & 17.5    & 2.90     & 46 & 0.12  & 97.17  & 0.22  & 11.0  & 153.0  & 15.6 & 0.75  \\
\hline
MDHM  & 17.5    & 2.90     & 39 & 0.12  & 114.31 & 0.77  & 8.0   & 141.8  & 15.2 & 0.74 \\
\hline
LDHM  & 17.5    & 2.90     & 33 & 0.12  & 134.49 & 0.50  & 8.7   & 133.8  & 14.1 & 0.73 \\
\hline
VLDHM & 17.5    & 2.90     & 23 & 0.12  & 194.34 & 0.83  & 6.6   & 112.5  & 10.8 & 0.63 \\
\hline
HDMM  & 0.269   & 0.0445   & 54 & 0.11  & 20.99  & 1.0   & 5.8   & 37.6   & 5.5  & 0.020 \\
\hline
MDMM  & 0.269   & 0.0445   & 45 & 0.11  & 24.69  & 0.75  & 7.0   & 35.7   & 5.4  & 0.019 \\
\hline
LDMM  & 0.269   & 0.0445   & 38 & 0.11  & 29.05  & 0.58  & 8.1   & 33.3   & 4.8  & 0.019 \\
\hline
VLDMM & 0.269   & 0.0445   & 26 & 0.11  & 41.98  & 0.15  & 11.8  & 28.3   & 4.7  & 0.017 \\
\hline
HDLM  & 0.00417 & 0.000691 & 63 & 0.09  & 4.48   & 0.36  & 9.7   & 9.2    & 2.3  & 0.00015 \\
\hline
MDLM  & 0.00417 & 0.000691 & 53 & 0.09  & 5.27   & 0.22  & 11.0  & 10.0   & 2.1  & 0.00014 \\
\hline
LDLM  & 0.00417 & 0.000691 & 45 & 0.09  & 6.20   & 0.05  & 13.0  & 11.8   & 2.0  & 0.00014 \\
\hline
VLDLM & 0.00417 & 0.000691 & 31 & 0.09  & 8.96   & 0.0   & 13.7  & 10.5   & 2.5  & 0.00010 \\
\hline
\end{tabular}
\label{tab1}
\end{table*}

%%%%%%%%%%%%%%%%%%%%%%%%%%%%%%Table 2
\begin{table*}%%%%%%%%%%[h!]
\caption{The ancillary models. The meaning of the symbols is as follows: Model is the two-letter string identifying the model according to the mass: MM for intermediate mass galaxy $2.69 \times 10^{11}M_\odot$ and LM for the low mass case $4.17 \times 10^{9}M_\odot$;  $f_\delta$ is the multiplicative factor of the initial over-density, in other words the starting over-density of the simulation is a factor $f_\delta $ higher that the  standard over-density  currently assumed for the reference model of the same mass; $\epsilon_{sf}$ is the dimensionless efficiency of the star
formation rate, the symbol $\epsilon_{sf}(Z)$ means that the efficiency is supposed to increase from $\epsilon_{sf}=0.1$ for $Z$=0.0001 to $\epsilon_{sf} = 1$ for $Z$=0.01 (close to the solar value). All other symbols have the same meaning as in Table \ref{tab1}.  } \centering
\begin{tabular}{|r|r|l |l|l|r  |l|l|l |l|l|l |l|}
\hline
Model &$f_\delta$ &$\epsilon_{sf}$   &$M_{t}$& $M_{g,i}$  & $z_{i}$      & $[\delta\rho-1]_{i}$ & $R_{i}$ & $z_{f}$ & $t_{f}$ & $R_{vir}$ & $R_{1/2}$ & $M_{s}$  \\
\hline
MM    &  20     &    1    & 0.269 & 0.0445     & 181 & 0.18 & 6.3  & 62.0  & 0.04   &  1.8   & 0.63  & 0.010\\
\hline
MM    &  20     &    1    & 0.269 & 0.0445     & 181 & 0.18 & 6.3  & 49.0  & 0.05   &  2.6   & 0.79  & 0.020 \\
\hline
MM    &  20     &    1    & 0.269 & 0.0445     & 181 & 0.18 & 6.3  & 30.4  & 0.10   &  4.9   & 1.26  & 0.054 \\
\hline
MM    &  20     &    0.1  & 0.269 & 0.0445     & 181 & 0.18 & 6.3  & 56.0   & 0.04  &  2.1   & 0.08  & 0.0050 \\
\hline
MM    &  20     &    0.1  & 0.269 & 0.0445     & 181 & 0.18 & 6.3  & 49.0  & 0.05   &  2.6   & 0.08  & 0.0074  \\
\hline
MM    &  20     &    $\epsilon_{sf}(Z)$  & 0.269 & 0.0445     & 181 & 0.18 & 6.3  & 59.0  & 0.04   &  2.0   & 0.07  & 0.0036 \\
\hline
MM    &  20     &    $\epsilon_{sf}(Z)$  & 0.269 & 0.0445     & 181 & 0.18 & 6.3  & 46.0  & 0.05   &  2.9   & 0.09  & 0.0083 \\
\hline
MM    &  15     &    1    & 0.269 & 0.0445     & 140 & 0.18 & 8.2  & 30.4  & 0.10   &  4.5   & 1.08  & 0.026 \\
\hline
MM    &  12     &    1    & 0.269 & 0.0445     & 117 & 0.17 & 9.8  & 30.4  & 0.10   &  4.1   & 0.92  & 0.011 \\
\hline
MM    &  12     &    1    & 0.269 & 0.0445     & 117 & 0.17 & 9.8  & 18.8  & 0.20   &  7.4   & 0.98  & 0.015 \\
\hline
MM    &  12     & $\epsilon_{sf}(Z)$  & 0.269 & 0.0445     & 117 & 0.17 & 9.8  & 36.0  & 0.08   &  3.1   & 0.07  & 0.0032 \\
\hline
MM    &  12     & $\epsilon_{sf}(Z)$  & 0.269 & 0.0445     & 117 & 0.17 & 9.8  & 30.4  & 0.10   &  4.2   & 0.10  & 0.00575 \\
\hline
MM    &  5      &    1    & 0.269 & 0.0445     & 68  & 0.14 & 16.7 & 7.0   & 0.70   &  19.2  & 5.75  & 0.019 \\
\hline
MM    &  5      &    1    & 0.269 & 0.0445     & 68  & 0.14 & 16.7 & 4.1   & 1.50   &  34.3  & 6.76  & 0.057 \\
\hline
LM    &  20     &    1    & 0.00417 & 0.000691 & 112 & 0.16 & 2.6  & 25.3  & 0.13   &  1.2   & 0.32  & 0.00016 \\
\hline
LM    & 20      &    1    & 0.00417 & 0.000691 & 112 & 0.16 & 2.6  & 18.8  & 0.20   &  1.8   & 0.31  & 0.00017 \\
\hline
LM    & 20      &    0.1  & 0.00417 & 0.000691 & 112 & 0.16 & 2.6  &  25.4  & 0.13  &  1.3   & 0.02  & 0.000091 \\
\hline
\end{tabular}
\label{tab2}
\end{table*}
\label{lastpage}

%================== End  APPENDIX ===========================%
\end{document}